\documentclass[a4paper,twoside]{llncs}

\let\fns\footnotesize

\usepackage{times}
\usepackage{url}
\usepackage{graphicx}
\usepackage{version}
\usepackage{alltt}
\usepackage{multirow}
\usepackage{bigdelim}       
\usepackage{amstext}
\usepackage{amsmath}
\usepackage{amssymb}
\usepackage{color}
\usepackage{textcomp}
\usepackage{rotating}
\usepackage{comment}

\usepackage{subfigure}
\usepackage[small]{caption}
\usepackage{floatflt}
\usepackage{enumerate}        

\usepackage{mathrsfs}         

\usepackage{bussproofs}       
\usepackage{dsfont}           
\usepackage{marginnote}

\usepackage{changepage}

\usepackage{longtable}
\usepackage{colortbl}

\usepackage{graphicx}
\usepackage{type1cm}
\usepackage{eso-pic}
\usepackage{color}
\makeatletter
\def\thewatermark{\begin{tabular}{c}THIS SECTION\\NOT FOR PUBLICATION\end{tabular}}
\def\watermarkangle{45}
\def\watermarklightness{0.9}
\def\watermarkfontsize{2cm}
\def\watermark{
    \AddToShipoutPicture{%
            \setlength{\@tempdimb}{.5\paperwidth}%
            \setlength{\@tempdimc}{.5\paperheight}%
            \setlength{\unitlength}{1pt}%
            \put(\strip@pt\@tempdimb,\strip@pt\@tempdimc){%
        \makebox(-20,100){\hss\rotatebox{\watermarkangle}{\textcolor[gray]{\watermarklightness}%
                 {\fontsize{\watermarkfontsize}{\watermarkfontsize}\selectfont{\thewatermark}}}\hss}%
        }%
    }%
}

\makeatother

\usepackage{skull} 

\newcommand{\reg}[1]{{\bf #1}}
\newcommand{\opcode}[1]{{\textbf{#1}}}

\newcommand{\type}[1]{{\textbf{#1}}}
\newcommand{\typevar}[1]{{\textbf{\emph{#1}}}}

\def\fract#1#2[#3]{
\frac{
\mbox{\fns$#1$}
}{
\mbox{\fns$#2$}
}
\mbox{\fns[$#3$]}
}
\def\pre[#1]#2{[#1]\mathop{{#2}}}

\def\to{{\rightarrow\kern0.5pt}}

\def\plus{\mathop{\mbox{\bf+}}}
\def\equals{\mathop{\mbox{\bf=}}}
\def\R{{\mathcal R}}
\def\K{{\mathcal K}}
\def\H{{\mathcal H}}

\newenvironment{leftprogram}{%
 \begin{flushleft}\begin{footnotesize}%
 \begin{alltt}
}{%
 \end{alltt}
 \end{footnotesize}\end{flushleft}%
}

\makeatletter
\newenvironment{textbox}[1][]{%
  \def\@captype{box}%
    \par\nobreak\vspace{-2ex}\begin{center}\nobreak\begin{framebox}%
}{%
  \end{framebox}\par\nobreak\end{center}\vspace{-0.5ex}%
}
\newcounter{textbox}

\def\newdef#1#2{%
    \expandafter\@ifdefinable\csname #1\endcsname
        {\@definecounter{#1}%
         \expandafter\xdef\csname the#1\endcsname{\@thmcounter{#1}}%
         \global\@namedef{#1}{\@defthm{#1}{#2}}%
         \global\@namedef{end#1}{\@endtheorem}%
    }%
}
\def\@defthm#1#2{%
    \refstepcounter{#1}%
    \@ifnextchar[{\@ydefthm{#1}{#2}}{\@xdefthm{#1}{#2}}%
}
\def\@xdefthm#1#2{%
    \@begindef{#2}{\csname the#1\endcsname}%
    \ignorespaces
}
\def\@ydefthm#1#2[#3]{%
    \trivlist
    \item[%
        \hskip 10\p@
        \hskip \labelsep
        {\it #2%
         \saveb@x\@tempboxa{#3}
         \ifdim \wd\@tempboxa>\z@
            \ \box\@tempboxa
         \fi.%
        }]%
    \ignorespaces
}
\def\@begindef#1#2{%
    \trivlist
    \item[%
        \hskip 10\p@
        \hskip \labelsep
        {\it #1\ \rm #2.}%
    ]%
}
\makeatother

\newdef{fact}{Fact}

\newenvironment{notation}[1][]{%
  \emph{Notation}.
  }{%
}

\renewcommand{\appendix}{\par
\section*{APPENDIX -- {\em NOT FOR PUBLICATION\/}}
\setcounter{section}{0}
 \setcounter{subsection}{0}
  \def\thesection{\Alph{section}} }

\excludecomment{anonymous}\includecomment{standard}
\excludecomment{withappendix}\includecomment{withoutappendix}

\begin{anonymous}

\def\citeBB#1{\ifnum#1=11 \expandafter\mycite{BB11A}\else\ifnum#1=94 \expandafter\mycite{BB94A}\else\ifnum#1=13 \expandafter\mycite{BB13A}\fi\fi\fi}

\end{anonymous}
\begin{standard}

\def\citeBB#1{\ifnum#1=11 \expandafter\mycite{BB11}\else\ifnum#1=94 \expandafter\mycite{BB94}\else\ifnum#1=13 \expandafter\mycite{BB13}\fi\fi\fi}

\def\andciteBBtwo#1,{BB#1,\andciteBBthree}
\def\andciteBBthree#1,{BB#1}

\end{standard}

\def\mycite#1{\cite{#1}}

\def\paragraph#1{{\bf#1}}

\begin{standard}
\def\authorA{Peter~T.~Breuer}
\def\authorAemail{ptb@cs.bham.ac.uk}

\def\authorAaddr{Department of Computer Science, University of Birmingham, UK}
\def\authorB{Jonathan~P.~Bowen}
\def\authorBemail{jonathan.bowen@lsbu.ac.uk}

\def\authorBaddr{Department of Informatics, London South Bank University, UK}
\end{standard}

\begin{anonymous}
\def\authorA{Author~1}
\def\authorAemail{Author 1 email}

\def\authorAaddr{Author 1 address}
\def\authorB{Author~2}
\def\authorBemail{Author 2 email}

\def\authorBaddr{Author 2 address}
\end{anonymous}

\def\nedots{{.^{.^.}\kern-5pt}}

\begin{document}

\edef\marginnotetextwidth{\the\textwidth}

\title{Certifying Machine Code Safe from Hardware Aliasing}
\subtitle{RISC is not necessarily risky}

\author{
\authorA\inst{1}
\and
\authorB\inst{2}
}
\institute{
  \authorAaddr\\
  \email{\authorAemail}
\and
  \authorBaddr\\
  \email{\authorBemail}
}

\maketitle

\vspace{-4ex}
\begin{abstract}
Sometimes machine code turns out to be a better target for verification
than source code.  RISC machine code is especially advantaged with
respect to source code in this regard because it has only two
instructions that access memory.  That architecture forms the basis here
for an inference system that can prove machine code safe against
`hardware aliasing', an effect that occurs in embedded systems.
There are programming memes that ensure code is safe from hardware
aliasing, but we want to certify that a given machine code is provably
safe.

\end{abstract}


\pagestyle{plain}

\section{Introduction}
\label{sec:Introduction}

In a computer system, `software' aliasing occurs when different
logical addresses simultaneously or sporadically reference the same
physical location in memory.  We are all familiar with it and think
nothing of it, because the same physical memory is nowadays reused
millisecond by millisecond for different user-space processes with
different addressing maps, and we expect the operating system kernel to
weave the necessary illusion of separation.  The kernel programmer has
to be aware that different logical addresses from different or even the
same user-space process may alias the same physical location, but the
application programmer may proceed unawares.

We are interested in a converse situation, called `hardware'
aliasing, where different physical locations in memory are
sporadically bound to the same logical address. 
If software aliasing is
likened to one slave at the beck of two masters, hardware aliasing
is like identical twins slaved to one master who cannot tell which is
which.  In this paper we will investigate the safety of
machine code in the light of hardware aliasing issues.

Aliasing has been studied before \mycite{sato1997speculative} and
is the subject of some patents
\mycite{fischer2002memory,wing1999method}.  There appears to be no
theoretical treatment published, although the subject is
broadly treated in most texts on computer architecture (see, for
example, Chapter 6 of \mycite{Barr98}) and is common lore in operating
systems kernel programming.  The `hardware' kind of aliasing
arises particularly in embedded systems where the arithmetic components
of the processor are insufficient to fill all the address lines.
Suppose, for example, that the memory has 64-bit addressing but the
processor only has 40-bit arithmetic.  The extra lines might be
grounded, or sent high, and this varies from platform to platform.  They may
be connected to 64-bit address registers in the processor, so their
values change from moment to moment as the register is filled.  In that
case, it is up to the software to set the `extra' bits reliably to zero,
or one, or some consistent value, in order that computing an address may
yield a consistent result.

We first encountered the phenomenon in the context of the KPU
\citeBB{13}, a general purpose `crypto-processor', i.e., a processor
that performs its computations in encrypted form in order to provide
security against observation and protection from malware.  Because real
encryptions are one-to-many, the result of the encrypted calculation of
the address $1+1$ will always mean `$2$' when decrypted, but may be
different from another encryption of $2$.  If the two different
{\em physical aliases} are used as addresses, then two different memory
cell contents are accessed and the result is chaotic.
The same effect occurs in the embedded system that has processor
arithmetic with fewer bits than there are address lines; add $1+1$ in
the processor and instead of $2$, $\rm 0xff01000000000002$ may be
returned.  If those two aliases of the arithmetic `$2$' are used as
addresses, they access different memory cells. The upshot is that
what is meant both times to be `$2$' accesses different locations
according to criteria beyond the programmer's control.



There are
programming memes that are successful in an aliasing environment:
if a pointer is needed again
in a routine, it must be copied exactly and saved for the next use;
when an array
or string element is accessed, 
the address must always be calculated in exactly the same way. 
But whatever the programmer says, the compiler may implement as it
prefers and ultimately it is the machine code that has to be checked in
order to be sure that aliasing is not a risk at run-time.
Indeed, in an embedded environment it is usual to find the programmer
writing in assembler precisely in order to control the machine code
emitted.  The Linux kernel consists of about $5\%$ hand-written assembly
code, for example (but rarely in segments of more than 10-15 lines each).
One of our long term objectives is to be able to boot a Linux kernel on an
embedded platform with aliasing, the KPU in particular. That requires
both modifying a compiler and checking the hand-written machine-level
code in the open source archive.

An inference system will be set out here that can guarantee a (RISC
\mycite{bowen1990formal,Pat85}) machine code program safe against
hardware aliasing as described.  The idea is to map a stack machine
onto the machine code.  We will reason about what assembly language
instructions for the stack machine do computationally.
Choosing an inference rule to apply to a machine code instruction is
equivalent to choosing a stack machine assembly language \cite{BB11} instruction
to which it disassembles \cite{BB93,BB94}.  The choice must be such that
a resulting proof tree is well-formed, and that acts as a guide.
The stack machine is aliasing-proof when operated within its intended
parameters so verifying alias-safety means verifying that the stack
machine assembly language code obtained by disassembly of the RISC
machine code does not cause the stack machine to overstep certain
bounds at run-time.
 
The RISC machine code we can check in this way is {\em ipso facto} restricted
to that which we can disassemble.  At the moment, that means code that
uses string or string-like data structures and arrays which do not
contain further pointers, and which uses machine code `jump and link' and `jump
register' instructions only for subroutine call and return respectively,
and in which subroutines make their own local frame and do not access
the caller's frame (arguments are passed to subroutines in registers).
These restrictions are not fundamental, but in any case there are no
functional limitations implied by them; one call convention is
functionally as good as another and data structures may always be laid
out flat, as they are in a relational DB.

Mistakes in disassembly are possible: if a `jump register' instruction,
for example, were in fact used to implement a computed goto and not a
subroutine return, it could still be treated as a subroutine return by
the verification, which would end prematurely, possibly missing an error
further along and returning a false negative.  A mistaken return as
just described would always fail verification in our system, but other
such situations are conceivable in principle.
So a human needs to check and certify that the proposed disassembly is
not wrongheaded.  The practice is not difficult because, as noted above,
hand-written machine code at a professional standard consists of short,
concise, commented segments.  The difficulty is that there is often a
great deal of it to be checked and humans tire easily. But our system
reduces the burden to checking the disassembly proposed by the
system against the comments in the code.

This paper is structured as follows: after an illustration of
programming against aliasing in Section~\ref{Sec:Getting it right:
programming memes} and a discussion of disassembly in
Section~\ref{sec:Disassembly},  code annotation is introduced in
sections \ref{Sec:annotation}, \ref{sec:Stack, string and array pointers}
and \ref{sec:Formal logic}, with a worked example in
Section~\ref{sec:Is any substantive program allowed?}.
Section~\ref{sec:Preventing unsafe memory access} argues that
code annotation gives rise to the formal assurance that
aliasing cannot occur.

\section{Programming memes}
\label{Sec:Getting it right: programming memes}

We model aliasing as being introduced when memory addresses are
calculated in different ways.  That model says that a memory address may
be {\em copied} exactly and used again without hazard, but if even $0$ is added
to it, then a different alias of the address may result, 
and reads from the new alias do not return data deposited at the old
alias of the address. Arithmetically the aliases are equivalent in the
processor; they will test as equal but they are not identical, and
using them as addresses shows that up.

\begin{floatingtable}[l]{
\begin{tabular}{@{}|@{~}l@{~}|@{~}l@{~}|@{}}
\hline
\hfill $\skull$\hfill  &\hfill  $\checkmark$\hfill \\[1ex]
\begin{minipage}[b]{0.20\textwidth}
\begin{leftprogram}
foo:\\
\quad            \\
\quad sp -{\bf=} 32         \\
\quad \mbox{\rm\dots {code} \dots}\\
\quad sp +{\bf=} 32\\
\quad return
\end{leftprogram}
\end{minipage}
&
\begin{minipage}[b]{0.20\textwidth}
\begin{leftprogram}
foo:\\
\quad gp ~{\bf=} sp\\
\quad sp -{\bf=} 32\\
\quad \mbox{\rm\dots {code} \dots}\\
\quad sp ~{\bf=} gp\\
\quad return
\end{leftprogram}
\end{minipage}
\\
\hline
\end{tabular}
}
\caption{Aliasing in function \emph{foo}.}
\label{tab:1}
\end{floatingtable}

That is particularly a problem for the way in which a compiler
-- or an assembly language programmer -- renders machine code
for the stack pointer movement around a function call.
Classically, a subroutine starts by decrementing the stack
pointer to make room on the stack for its local frame.  Just
before return, it increments the stack pointer back to its original value.
The pseudo-code is shown on the left in Table~\ref{tab:1}.
In an aliasing context, the attempt at arithmetically restoring the
pointer puts an alias of the intended
address in the \reg{sp} register, and the caller may receive
back a stack pointer that no longer points to the data. The
code on the right in Table~\ref{tab:1} works correctly;
it takes an extra register (\reg{gp}) and instruction, but the register
content may be moved to the stack and restored before return, avoiding
the loss of the slot.

\begin{floatingtable}[l]{
\begin{tabular}{@{}|@{\,}l@{\,}|@{~}l@{~}|@{~}l@{~}|@{~}l@{~}|@{}}
\hline
\em string &
\hfill $\skull$\hfill  &
\hfill $\skull$\hfill  &
\hfill  $\checkmark$\hfill \\[1ex]
\em array &
\hfill $\checkmark$\hfill  &
\hfill $\skull$\hfill  &
\hfill  $\skull$\hfill \\[1ex]
&
\begin{minipage}[b]{0.155\textwidth}
\begin{leftprogram}
\quad\\
\quad x = s[2]
\end{leftprogram}
\end{minipage}
&
\begin{minipage}[b]{0.155\textwidth}
\begin{leftprogram}
\quad s+= 2\\
\quad x = *s
\end{leftprogram}
\end{minipage}
&
\begin{minipage}[b]{0.155\textwidth}
\begin{leftprogram}
\quad s++; s++\\
\quad x = *s
\end{leftprogram}
\end{minipage}
\\
\hline
\end{tabular}
}
\caption{Aliasing while accessing a string or array.  }
\label{tab:2}
\end{floatingtable}
Strings and arrays are also problematic in an aliasing environment
because different calculations for the address of the same element
cause aliasing.
To avoid it, the strategy we will follow is
that elements of `string-like' structures will be accessed by
incrementing the base address in constant steps (see the pseudo-code 
at right in Table~\ref{tab:2}) and array elements will be accessed via a unique
offset from the array base address (see the pseudo-code at left in
Table~\ref{tab:2}).  This technique
ensures that there is only one calculation possible for the address of
each string element (it is $((s{+}1){+}1){+}0$ in Table~\ref{tab:2}) or array
element ($s{+}2$ in Table~\ref{tab:2}), so aliasing cannot occur.  The
middle code in Table~\ref{tab:2} gives address $(s{+}2){+}0$ which matches
exactly neither string nor array calculations. The
decision over whether to treat a memory area like a string or an array
depends on the mode of access to be used.


\section{Disassembly}
\label{sec:Disassembly}

\begin{table}[tb]
 \fbox{
  \begin{minipage}{0.97\textwidth}
   \begin{scriptsize}
A RISC machine code processor
consists of 32 (32-bit) integer registers $R$, a vector of
$2^{32}$ (32-bit) integer memory locations $M$, and the
program counter $p$. The latter gives the address of the current
instruction.  The {\reg{ra}} register is used to hold a subroutine
call return address.  Only two instructions, \opcode{sw}
and \opcode{lw}, access memory.
   \end{scriptsize}
   \begin{center}
    \scriptsize
    \begin{tabular}{@{}l|l|l@{}}
instruction &mnemonic & semantics\\
\hline&&\\[-2ex]
${\bf sw}~r_1~k(r_2)$     &store word &$M'=M\oplus\{R\,r_2\plus k\mapsto
R\,r_1\};~R'=R;~p'=p{+}4$\\
${\bf lw}~r_1~k(r_2)$     &load word      &$M'=M;~R'=R\oplus\{r_1
\mapsto M(R\,r_2\plus k)\};~p'=p{+}4$\\
${\bf move}~r_1~r_2$      &move/copy      &$M'=M;~R'=R\oplus\{r_1\mapsto
R\,r_2\};~p'=p{+}4$\\
${\bf li}~r_1~k$          &load immediate &$M'=M;~R'=R\oplus\{r_1\mapsto
k\};~p'=p{+}4$\\
${\bf addiu}~r_1~r_2~k$    &add immediate  &$M'=M;~R'=R\oplus\{r_1\mapsto
R\,r_2 \plus k\};~p'=p{+}4$\\
${\bf addu}~r_1~r_2~r_3$   &add variable   &$M'=M;~R'=R\oplus\{r_1\mapsto
R\,r_2 \plus R\,r_3\};~p'=p{+}4$\\
${\bf nand}~r_1~r_2~r_3$  &bitwise not-and&$M'=M;~R'=R\oplus\{r_1\mapsto
R\,r_2 \mathop{\overline{\mbox{\bf\&}}} R\,r_3\};~p'=p{+}4$\\
${\bf beq}~r_1~r_2~k$     &branch-if-equal&$M'=M;~R'=R;~{\bf
if}\,(R\,r_1\equals R\,r_2)~p'=k~{\bf else}~p'=p{+}4$\\
${\bf jal}~k$             &jump-and-link  &$M'=M;~R'=R\oplus\{{\bf
ra}\mapsto p{+}4\};~p'=k$\\
${\bf jr}~r$                &jump-register  &$M'=M;~R'=R;~p'=R\,r$\\
    \end{tabular}
   \end{center}
   \begin{scriptsize}
    \begin{notation}
$M\oplus\{a\mapsto v\}$ means the
vector $M$ overwritten at index $a$
with the value $v$; the processor arithmetic (bold font `$\plus$') is
distinguished from the instruction addressing arithmetic (light font
`$+$'); $r_1$, $r_2$ are register
names or indices; $k$ is a signed 16-bit integer; $x$ and $x'$ are
respectively initial and final value after the instruction has acted.
    \end{notation}
   \end{scriptsize}
  \end{minipage}
 }
 \vspace{2ex}
 \def\tablename{Box}
 \def\thetable{1}
 \addtocounter{table}{-1}
 \caption{RISC machine code instructions and their underlying semantics.}
 \label{Tab:3}
\end{table}

\begin{floatingtable}{
\kern-7pt\hbox to 0.5\textwidth {
\vbox to 1.05in{
\fbox{
\begin{minipage}{0.48\textwidth}
\medskip
\scriptsize
%
Say that the stack pointer $s$ is in the stack pointer
register \reg{sp} in the machine code processor.
A corresponding abstract stack machine state is a 4-tuple $(\R, \K, \H, p)$,
where $\R$ consists of the
$31$ registers excluding the stack pointer register, the stack $\K$
consists of the top part of memory above the stack pointer value $s$,
the heap $\H$ consists of the bottom part of memory below the
stack pointer, and the address $p$ is that of the current instruction.
\begin{align*}
\K \,k &= M (s \plus k) &s &= R\,\reg{sp},~ k \ge  0 \\
\R \,r &= R\,r        &r &\ne \reg{sp},~ r \in \{0, \dots 31\} \\
\H \,a &= M\,a       &a &<s
\end{align*}
The (hidden) stack pointer value $s$ is needed to recreate the machine code
processor state $(R,M,p)$ from the stack machine state $(\R, \K, \H,
p)$, so the latter is more abstract.
\end{minipage}
}
}
\hss
}
\kern2pt
}
\def\tablename{Box}
\def\thetable{2}
\addtocounter{table}{-1}
\caption{Relation of processor to stack machine.}
\label{tab:rel}
\end{floatingtable}

Nothing in the machine code indicates which register holds a
subroutine return address, and that affects wh\-ich machine code
instructions may be interpreted as a return from a subroutine call.
To deal with this and similar issues in an organised manner, we describe
rules of reasoning about programs both in terms of the machine code
instruction to which they apply and an assembly language instruction for
a more abstract {\em stack machine} that the machine code instruction
may be disassembled to and which we imagine the programmer is 
targeting.

The core RISC machine code instructions are listed in Box~\ref{Tab:3},
where their semantics are given as state-to-state transformations on the
three components of a RISC processor: 32 32-bit registers $R$, memory
$M$ and a 32-bit program counter $p$.  The corresponding abstract stack
machine is described in Box~\ref{tab:rel}.  The stack pointer address
$s$ in the machine code processor notionally divides memory $M$
into two components: stack $\K$ above and heap $\H$ below.
The stack machine manipulates the stack directly via instructions that
operate at the level of stack operations, and they are implemented in
the machine code processor via instructions that act explicitly on the
stack pointer.  No stack pointer is available in the abstract machine.
Its registers $\R$ consist of the set $R$ in the machine code processor
{\em minus} the register that contains the stack pointer, usually the
\reg{sp} register.  The program counter $p$ is the same in the abstract
stack machine as in the machine code processor, because instructions
correspond one-to-one between programs for each machine.  However, there
is usually a choice of more than one abstract stack machine instruction
that each machine code instruction could have been disassembled to, even
though only one is chosen.

\begin{table}[tb]
\caption{
Stack machine instructions: the $n$ are small integers,
the $r$ are register names or indices, and the $a$ are relative or
absolute addresses.
}
\label{tab:3}
\[
\begin{array}{lcll}
s&\mathop{{:}{:}{=}} &
                       \opcode{cspt}~$r$
                   ~|~ \opcode{cspf}~$r$
                   ~|~ \opcode{rspf}~$r$
                   ~|~ \opcode{push}~$n$
                     & \mbox{// stack pointer movement}
                \\
                   &|& \opcode{get}~$r$~$n$
                   ~|~ \opcode{put}~$r$~$n$
                   ~|~ \dots
                     & \mbox{// stack access}
                \\
                   &|& \opcode{newx}~$r$~$a$~$n$
                   ~|~ \opcode{stepx}~$r$~$n$
                   ~|~ \opcode{getx}~$r$~$n(r)$
                   ~|~ \opcode{putx}~$r$~$n(r)$
                   ~|~ \dots
                     & \mbox{// string operations}
                \\
                   &|& \opcode{newh}~$r$~$a$~$n$
                   ~|~ \opcode{lwfh}~$r$~$n(r)$
                   ~|~ \opcode{swfh}~$r$~$n(r)$
                   ~|~ \dots
                     & \mbox{// array operations}
                \\
                   &|& \opcode{gosub}~$a$
                   ~|~ \opcode{return}
                   ~|~ \opcode{goto}~$a$
                   ~|~ \opcode{ifnz}~$r$~$a$
                   ~|~ \dots
                     & \mbox{// control operations}
                \\
                   &|& \opcode{mov}~$r$~$r$
                   ~|~ \opcode{addaiu}~$r$~$r$~$n$
                   ~|~ \dots
                     & \mbox{// arithmetic operations}
                \\[-6ex]
\end{array}
\]
\end{table}

\begin{table}[tb]
\caption{
Machine code may be disassembled to one of several alternate assembly
language instructions for a stack machine.
}
\begin{center}
\begin{tabular}[t]{|l|l|}
\hline
machine code & assembly language \\[0.5ex]
\hline
\opcode{move} $r_1$ $r_2$ &
    \begin{tabular}{l}
    \opcode{cspt} $r_1$\\
    \opcode{cspf} $r_2$\\
    \opcode{rspf} $r_2$\\
    \opcode{mov}  $r_1$ $r_2$
    \end{tabular}\\
\hline
\opcode{addiu} $r$ $r$ $n$ &
    \begin{tabular}{l}
    \opcode{push} $\mbox{-}n$\\
    \opcode{stepx} $r$ $n$\\
    \opcode{addaiu} $r$ $r$ $n$
    \end{tabular}\\
\hline
\opcode{lw} $r_1$ $n(r_2)$ &
    \begin{tabular}{l}
    \opcode{get} $r_1$ $n$\\
    \opcode{lwfh} $r_1$ $n(r_2)$\\
    \opcode{getx} $r_1$ $n(r_2)$
    \end{tabular}\\
\hline
\opcode{sw} $r_1$ $n(r_2)$ &
    \begin{tabular}{l}
    \opcode{put} $r_1$ $n$\\
    \opcode{swfh} $r_1$ $n(r_2)$\\
    \opcode{putx} $r_1$ $n(r_2)$
    \end{tabular}\\
\hline
\end{tabular}
\qquad
\begin{tabular}[t]{|l|l|}
\hline
machine code & assembly language \\[0.5ex]
\hline
\opcode{lb} $r_1$ $n(r_2)$ &
    \begin{tabular}{l}
    \opcode{getb} $r_1$ $n$\\
    \opcode{lbfh} $r_1$ $n(r_2)$\\
    \opcode{getbx} $r_1$ $n(r_2)$
    \end{tabular}\\[0.5ex]
\hline
\opcode{sb} $r_1$ $n(r_2)$ &
    \begin{tabular}{l}
    \opcode{putb} $r_1$ $n$\\
    \opcode{sbth} $r_1$ $n(r_2)$\\
    \opcode{putbx} $r_1$ $n(r_2)$
    \end{tabular}\\[0.5ex]
\hline
\opcode{jal}~$a$ &
    \begin{tabular}{l}
    \opcode{gosub}~$a$
    \end{tabular}\\[0.5ex]
\hline
\opcode{jr}~$r$ &
    \begin{tabular}{l}
    \opcode{return}
    \end{tabular}\\[0.5ex]
\hline
\opcode{j}~$a$ &
    \begin{tabular}{l}
    \opcode{goto}~$a$
    \end{tabular}\\[0.45ex]
\hline
\opcode{li}~$r$ $a$&
    \begin{tabular}{l}
    \opcode{newx}~$r$~$a$~$n$\\
    \opcode{newh}~$r$~$a$~$n$
    \end{tabular}\\[0.5ex]
\hline
\opcode{bnez}~$r$ $a$&
    \begin{tabular}{l}
    \opcode{ifnz}~$r$~$a$
    \end{tabular}\\[0.5ex]
\hline
\end{tabular}
\label{tab:4}
\end{center}
\vspace{-6ex}
\end{table}%


For example, several different stack machine instructions may all be
thought of as manipulating the hidden stack pointer, register \reg{sp}
in the machine code processor, and they all are implemented as a
\opcode{move} (`copy') machine code instruction. Thus the \opcode{move}
instruction disassembles to one of several stack machine instructions
as follows:

\medskip

\begin{enumerate}
\item
The \opcode{cspt}~$r_1$ (`copy stack pointer to') instruction
saves a copy of the stack pointer in register
$r_1$. It corresponds to the \opcode{move}~$r_1$~\reg{sp} machine code
processor instruction.

\item
The \opcode{cspf}~$r_1$ (`copy stack pointer
from') instruction {\em refreshes} the stack pointer
from a copy in $r_1$ that has the same value and was saved
earlier (we will not explore here the reasons why a compiler might issue
such a `refresh' instruction). It corresponds to the
\opcode{move}~\reg{sp}~$r_1$ machine code instruction.

\item
The \opcode{rspf}~$r_1$ (`restore stack
pointer from') instruction returns the stack pointer to a
value that it held previously by copying an old saved value from $r_1$.
It also corresponds to \opcode{move}~\reg{sp}~$r_1$.
\end{enumerate}
A fourth disassembly of the machine code \opcode{move} instruction,
to the stack machine \opcode{mov} instruction, encompasses the case when
the stack pointer is not involved at all; it does a straight copy of a
word from one register to another at the stack machine level.  The full
set of stack machine instructions is listed in Table~\ref{tab:3}, and
their correspondence with RISC machine code instructions is shown in
Table~\ref{tab:4}.

We will not work through all the instructions and disassembly options
in detail here, but note the important \opcode{push}~$n$
instruction in the stack machine, which can be thought of as decrementing
the hidden stack
pointer by $n$, extending the stack downwards. It corresponds to
the \opcode{addiu}~\reg{sp}~\reg{sp}~$m$ machine code instruction, with
$m=-n$.
Also, the stack machine instructions \opcode{put}~$r_1$~$n$ and
\opcode{get}~$r_1$~$n$ access the stack for a word at offset $n$ bytes,
and they correspond to the machine code \opcode{sw}~$r_1$~$n(\reg{sp})$ and
\opcode{lw}~$r_1$~$n(\reg{sp})$ instructions, respectively.

The very same machine code instructions may also be interpreted
as stack machine instructions that manipulate not the stack but either
a `string-like' object 
%
or an array. Strings/arrays
are read with \opcode{getx}/\opcode{lwfh} and written with
\opcode{putx}/\opcode{swth}.  Table~\ref{tab:4} shows that these are
implemented by \opcode{lw}/\opcode{sw} in the machine code processor,
applied to a base register $r_2\ne \reg{sp}$.
Stepping through a string is done with the \opcode{stepx} instruction in
the stack machine, which is implemented by \opcode{addiu} in the
machine code procesor. Introducing the address of a string/array in the
stack machine needs \opcode{newx}/\opcode{newh} 
and those are both implemented by the \opcode{li} (`load immediate')
instruction in the machine code processor.


There are also `b' (`byte-sized') versions of the \opcode{get},
\opcode{lwfh}, \opcode{getx} stack machine instructions
named \opcode{getb}, \opcode{lbfh}, \opcode{getbx} respectively. These 
are implemented by \opcode{lb} in the machine code
processor.
For \opcode{put}, \opcode{swth}, \opcode{putx} we have byte versions
\opcode{putb}, \opcode{sbth}, \opcode{putbx}.


\section{Introducing annotations and annotated types}
\label{Sec:annotation}


\begin{floatingtable}[l]{
\quad
\begin{minipage}[b]{0.26\textwidth}
\begin{leftprogram}
foo:\\
\quad move gp sp\\
\quad addiu sp sp -32\\
\quad \mbox{\rm\dots {code} \dots}\\
\quad move sp gp\\
\quad jr ra
\end{leftprogram}
\end{minipage}
}
\caption{Non-aliasing subroutine machine code.}
\label{tab:5}
\end{floatingtable}


Consider the `good' pseudo-code of Table~\ref{tab:1}
implemented as machine code and shown in Table~\ref{tab:5}. 
How do we show it is aliasing-safe? Our technique is to {\em annotate}
the code in a style akin to verification using Hoare logic, but the
annotation logic is based on the stack machine abstraction of what the
machine code
does. We begin with an annotation that says the \reg{sp} register is
bound to a particular {\em annotation type} on entry:
\[
\{\,\reg{sp} = \type{c}!0!4!8\,\}
\]
The `\type{c}' as base signifies a variable pointer value is in
register \reg{sp}.  It is the stack pointer value.  The `!0!4!8'
means that that particular value has been used
as the base address for writes to memory at offsets 0, 4 and 8 bytes
from it, respectively.

%
The first instruction in subroutine \emph{foo} copies the stack
pointer to register \reg{gp} and we infer that register \reg{gp}
also gets the `$\type{c}$' annotation, using a Hoare-triple-like
notation:
\[
  \{\,\reg{sp}^*= \type{c}!0!4!8\,\}
  ~
  \mbox{\tt move gp sp}
  ~
  \{\,\reg{sp}^*, \reg{gp}= \type{c}!0!4!8\,\}
\]
The stack pointer location (in the \reg{sp}
register) should always be indicated by an asterisk.

The arithmetic done by the next instruction destroys the offset
information.  It cannot yet be said that anything has been
written at some offset from the new address, which is 32
distant from the old only up to an arithmetic equivalence in the
processor: 
\[
  \{\,\reg{sp}^*, \reg{gp} = \type{c}!0!4!8\,\}
  ~
  \mbox{\tt addiu sp sp -32}
  ~
  \{\,\reg{gp} = \type{c}!0!4!8; ~ \reg{sp}^* = \type{c}\,\}
\]
Suppose the annotation on the \reg{gp} register is still valid
at the end of subroutine \emph{foo}, so the stack pointer
register is finally refreshed by the \opcode{move} instruction with the same
annotation as at the start:
\[
  \{\,\reg{sp}^*=\type{c};~\reg{gp}= \type{c}!0!4!8;\,\}
  ~
  \mbox{\tt move sp gp}
  ~
  \{\,\reg{sp}^*, \reg{gp} = \type{c}!0!4!8\,\}
\]
The return (\opcode{jr}~\reg{ra}) instruction does not change these
annotations.  So the calling code has returned as stack pointer a value
that is annotated as having had values saved at offsets 0, 4, 8 from it,
and the caller can rely on accessing data stored at those offsets.
That does not guarantee that the {\em same} value of the
stack pointer is returned to the caller, however.
It will be shown below how this system of annotations may be coaxed into
providing stronger guarantees.


\medskip
\medskip

\section{Types for stack, string and array pointers}
\label{sec:Stack, string and array pointers}

The annotation discussed above is not complete.  The {\em size} in
bytes of the local stack frame needs to be recorded by following the
`\type{c}' with the frame size as a superscript.  Suppose that on entry there
is a local stack frame of size 12 words, or 48 bytes.  Then here is the
same annotation with superscripts on, written as a derivation in which
the appropriate disassembly of each machine code instruction is written
to the right of the machine code as the `justification' for the
derivation: 
\begin{center}
\begin{minipage}{0.99\textwidth}
\begin{prooftree}
\AxiomC{\hbox to 1.7in{\{$\reg{sp}^* = \type{c}^{48}!0!4!8$\}\hfil}}
\RightLabel{\hbox to 0.95in{\opcode{move} gp sp\hfil}/~\opcode{cspt} gp}
\UnaryInfC{\hbox to 1.7in{\{$\reg{sp}^*,\reg{gp} = \type{c}^{48}!0!4!8$\}\hfil}}
\RightLabel{\hbox to 0.95in{\opcode{addiu} sp sp -32\hfil}/~\opcode{push} 32}
\UnaryInfC{\hbox to 1.7in{\{$ \reg{sp}^*=\type{c}^{32^{48}};~\reg{gp}=\type{c}^{48}!0!4!8$\}\hfil}}
\noLine
\UnaryInfC{\hbox to 1.7in{\hfil\vdots\hfil}}
\noLine
\UnaryInfC{\hbox to 1.7in{\{$\reg{sp}^*=\type{c}^{32^{48}};~ \reg{gp}=\type{c}^{48}!0!4!8$\}\hfil}}
\RightLabel{\hbox to 0.95in{\opcode{move} sp gp\hfil}/~\opcode{rspf} gp}
\UnaryInfC{\hbox to 1.7in{\{$\reg{sp}^*, \reg{gp}=\type{c}^{48}!0!4!8$\}\hfil}}
\end{prooftree}
\end{minipage}
\end{center}
The \opcode{push}~32 abstract stack machine instruction makes a {\em new}
local stack frame of $8$ words or $32$ bytes. It does not
increase the size of the current frame.
Accordingly, the $32$ `pushes up' the $48$ in the annotation so that
$32^{48}$ is shown.  This makes the size of the previous
stack frame available to the annotation logic.

%


\begin{floatingtable}{
 \fbox{
  \begin{minipage}{0.4\textwidth}
   \begin{scriptsize}
Annotations $a$ assert a binding of registers $r$ or
stack slots $(n)$ to an {\em annotated type} $t$. One of the
register names may be starred to indicate the stack pointer
position.  A type is either `uncalculated', \type{u}, or
`calculated', \type{c}.  Either may be decorated with `$!n$'
annotations indicating historical writes at that offset from the
typed value when used as an address.  A \type{c} base type may
also be superscripted by a `tower' of natural numbers $n$
denoting `frame sizes' (see text), while  a \type{u} base type
may have a single superscript (also denoting size). We also use
$\ddot{1}$ for a tower $1^{1^\nedots}$ of undetermined extent
and a single repeated size.  Also, formal type variables
$\typevar{x}$, $\typevar{y}$, etc are valid stand-ins for
annotated types, and formal `set of offsets variables'
$\typevar{X}$, $\typevar{Y}$, etc are valid stand-ins for sets
of offsets.
   \end{scriptsize}
\[
\begin{array}{lcl}
a&\mathop{{:}{:}{=}} &
r^{[\text{\bf*}]}\text{\bf,}\,\dots\text{\bf,}\,\text{\bf(}n\text{\bf)}\text{\bf,}\,\dots
\,\text{\bf=}\,t\text{\bf;}\,\dots\\
t&\mathop{{:}{:}{=}} & 
                     \type{c}^{[n^\nedots\kern3pt]}\text{\bf!}n\text{\bf!}\dots
                 ~|~ \type{u}^{[n]}\text{\bf!}n\text{\bf!}\dots
                 \\[-4ex]
\end{array}
\]
  \vspace{1ex}
  \end{minipage}
 }
}
 \vspace{1ex}
 \def\tablename{Box}
 \def\thetable{3}
 \addtocounter{table}{-1}
 \caption{Syntax of annotations and types.}
 \label{tab:ann}
\end{floatingtable}

A different disassembly of \opcode{addaiu}~$r$~$r$~$n$ is required
when $r$ contains a string pointer, not the stack pointer, which means
that register $r$ lacks the asterisk in the annotation.
The disassembly as
a step along a string is written \opcode{stepx}~$r$~$n$, and 
requires $n$ to be positive. In this case, the string pointer in  $r$ 
will be annotated with the type
\[
\type{c}^{\ddot{1}}
\]
meaning that it is a `calculatable' value that may be altered by adding
$1$ to it repeatedly.
The form $\type{c}^{\ddot{1}}$ hints that a string is regarded
as a stack $\type{c}^{1^\nedots}$ that starts `pre-charged' with an
indefinite number of frames of 1 byte each, which one may step
up through by `popping the stack' one frame, and one byte, at a time.
So annotation types may be either like
$\type{c}^{32^{48}}$ or $\type{c}^{\ddot{1}}$ and these may be followed by
offsets $!0!4!8!\dots$.  There is just one more base form,
described below, completing the list in Box~\ref{tab:ann}.

The RISC instruction \opcode{lw}~$r_1$~$n(r_2)$ is also disassembled
differently according to the annotated type in $r_2$. As \opcode{get}~$r_1$~$n$
it retrieves a value previously stored at offset $n$ in the stack, when
$n\ge 0$ and $r_2$ is the stack pointer register.  As
\opcode{lwfh}~$r_1$~$n(r_2)$ it retrieves an element in an array from the
{\em heap} area.  In that case, $r_2$ will be annotated
\[
\type{u}^m
\]
meaning an `unmodifiable' pointer to an array of size $m$ bytes, and
$m-4\ge n\ge 0$.  A third possibility is dissassembly as retrieval
from a string-like object in the heap, when, as
$\opcode{getx}~r_1~n(r_2)$,  register $r_2$ will
have a `string-like' annotation of the form $\type{c}^{\ddot{m}}$,
meaning that it must be stepped through in increments of $m$ bytes.

Similarly the RISC \opcode{sw}~$r_1$~$n(r_2)$ instruction can be
dissassembled as \opcode{put}~$r_1$~$n$ of a value at offset $n$ to the stack,
or \opcode{swth}~$r_1$~$n(r_2)$ to an array or $\opcode{putx}~r_1~n(r_2)$ to a
string, depending on the type bound to register $r_2$.
These register types drive the disassembly.
\begin{table}[t]
\caption{Possible disassemblies of machine code instructions as
constrained by the stack pointer register location changes
(SP$\leftarrow$SP) or absence ($\times$), and changes to the stack
content (`delta').
}
\medskip
\subtable{
\begin{tabular}{|l||c|c||l|}
\hline
move $r_1$ $r_2$~~~~~~&     $r_1$&$r_2$     & stack delta\\
\hline
\hline
rspf $r_2$    &     SP $\circlearrowright$  & $\times$ & yes\\
\hline
cspf $r_2$    &     SP $\circlearrowright$  & $\times$ & no\\
\hline
cspt $r_1$    &   $\times$     &SP $\circlearrowright$    & no \\
\hline
mspt $r_1$    &\multicolumn{2}{c||}{SP$\longleftarrow$SP} &   no\\
\hline
mov $r_1$ $r_2$&    $\times$   &   $\times$  &      no\\
\hline
\end{tabular}
}
\hfill
\subtable{
\begin{tabular}{|l||c|c||l|}
\hline
addiu $r_1$ $r_2$ $m$&  $r_1$&$r_2$  &  stack delta\\
\hline
\hline
step $r$ $m$ &      \multicolumn{2}{c||}{$\times$}&   no\\
\hline
stepto $r_1$ $r_2$ $m$ &      ~~ $\times$   ~~&  ~~$\times$ ~~&   no\\
\hline
push  ${-}m$     &     \multicolumn{2}{c||}{SP $\circlearrowright$ }&   yes\\
\hline
pushto $r_1$ ${-}m$ &    \multicolumn{2}{c||}{~SP$\longleftarrow$SP~~}&  yes\\
\hline
addaiu $r_1$ $r_2$ $m$&    ~~ $\times$   ~~&  ~~$\times$ ~~&       no\\
\hline
\end{tabular}
}
\begin{center}
~\begin{tabular}{|l||c|c||l|}
\hline
lw $r_1$ $m$($r_2$) &  $r_1$ & $r_2$ & stack delta\\
\hline
\hline
get $r_1$ $m$                 & ~~~$\times$~~~     & SP $\circlearrowright$ & no \\
\hline
lwfh $r_1$ $m$($r_2$)&    ~~~$\times$~~~   & ~~~$\times$~~~ & no    \\
\hline
getx $r_1$ $m$($r_2$)&    ~~~$\times$~~~   & ~~~$\times$~~~ & no    \\
\hline
\end{tabular}
\hfill
\begin{tabular}{|l||c|c||l|}
\hline
    \strut sw $r_1$ $m$($r_2$) &   $r_1$  &  $r_2$  & stack delta\\
\hline
\hline
    put $r_1$ $m$             & ~~~$\times$~~~     & SP $\circlearrowright$ & no\\
\hline
    \strut swth $r_1$ $m$($r_2$)&  ~~~$\times$~~~     &  ~~~$\times$~~~ & no   \\
\hline
    \strut putx $r_1$ $m$($r_2$)&  ~~~$\times$~~~     &  ~~~$\times$~~~ & no   \\
\hline
\end{tabular}~~
\end{center}
\vspace{-2ex}
\label{tab:6}
\end{table}

\section{Formal logic}
\label{sec:Formal logic}

We can now write down formal rules for the logic of annotations
introduced informally in
the `derivation' laid out in the previous section. Readers who would
prefer to see a worked example first should jump directly to
Section~\ref{sec:Is any substantive program allowed?}.

We start with a list of so-called `small-step' program annotations
justified by individual stack machine instructions, each the disassembly
of a machine code instruction.  The small-step rules relate the annotation
before each machine code instruction to the annotation after.
Table~\ref{tab:6} helps to reduce {\em a priori} the number of possible
disassemblies for each machine code instruction, but in principle
disassembly to stack machine code does not have to be done first, but
can be left till the last possible moment during the annotation process,
as each dissassembly choice corresponds to the application of a
different rule of inference about which annotation comes next.  If the
corresponding inference rule may not be applied, then that disassembly
choice is impossible.

Here is how to read Table~\ref{tab:7}.  Firstly, `offsets variables'
$\typevar{X}$, $\typevar{Y}$, etc, stand in for sets of offset
annotations `$!k$'.  For example, the $\opcode{put}~\reg{gp}~4$
instruction is expected to start with a prior annotation pattern
$\reg{sp}^*=\type{c}^{f}!\typevar{X}$ for the stack pointer register.
Secondly, the stack pointer register is indicated by
an asterisk. Thirdly, $f$ in the table stands for
some particular stack frame tower of integers;  it is not a variable,
being always some constant in any patrticular instance.  In the case of the
$\opcode{put}~\reg{gp}~4$ instruction, $f$ must start with some
particular number at least 8 in size, in order to accommodate the 4-byte
word written at offset 4 bytes within the local stack frame.  Just `$8$'
on its own would do for $f$ here. Lastly, `type variables' \typevar{x},
\typevar{y}, etc, where they appear, stand in for full types.

The table relates annotations before and after each instruction. So,
in the case of the $\opcode{put}~\reg{gp}~4$ instruction, if the prior
annotation for the stack pointer register is
$\reg{sp}^*=\type{c}^{f}!\typevar{X}$, then the post annotation is
$\reg{sp}^*=\type{c}^{f}!4!\typevar{X}$, meaning that 4 is one of the
offsets at which a write has been made.  It may be that 4 is also a
member of the set denoted by \typevar{X} (which may contain other
offsets too), or it may be not in \typevar{X}.  That is not decided by
the formula, which merely says that whatever other offsets there are in
the annotation, `4' is put there by this instruction.  At any rate, 
the annotation pattern for the $\opcode{put}~\reg{gp}~4$ instruction is:
\[
\{ \dots;\reg{sp}^*=\type{c}^{f}!\typevar{X};\dots \}
~\opcode{put}~\reg{gp}~4~
\{ \dots;\reg{sp}^*=\type{c}^{f}!4!\typevar{X};\dots \}
\]
and considering the effect on the \reg{gp} register (which may be
supposed to have the type denoted by the formal type variable
\typevar{x} initially) and the stack slot denoted by `(4)' gives
\[
\{ \reg{gp}{=}\typevar{x};\reg{sp}^*{=}\type{c}^{f}!\typevar{X} \}
~\opcode{put}~\reg{gp}~4~
\{ \reg{sp}^*{=}\type{c}^{f}!4!\typevar{X};\reg{gp}{,}(4){=}\typevar{x} \}
\]
because whatever the description \typevar{x} of the data in
register \reg{gp} before the instruction runs, since the data is
transferred to stack slot `($4$)', the latter gains the same
description.  Generalising 
the stack offset `$4$' back to $n$, and generalising
registers \reg{gp} and \reg{sp} to $r_1$ and $r_2$ respectively, one
obtains exactly the small-step signature listed for instruction
$\opcode{put}~r_1~n$.  Registers whose
annotations are not mentioned in this signature have bindings that
are unaffected by the instruction.



\begin{table}[tb]
\caption{
`Small-step' annotations on assembly instructions.
}
\label{tab:7}
\[
\begin{array}{@{}rcl@{~~}p{2in}@{}}
 \{~ \}
&\opcode{newx}~r~n &
\{ r\,\,{=}\type{c}^{\ddot{n}}!\typevar{X}
      \}
       &// \small Set reg. $r$  content\\
 \{r_1{=}\type{c}^{f_1}!\typevar{Y}{;\,}r_2{=}\type{u}^{f_2}!\typevar{X}
      \}
&\opcode{putx}~r_1~n(r_2) &
\{
 r_1{=}\type{c}^{f_1}!\typevar{Y}
 {;}\,
 r_2{=}\type{u}^{f_2}!n!\typevar{X}\}
       &// \small Store word to string\\
 \{
 r_2{=}\type{u}^{f}!n!\typevar{X}
 \}
&\opcode{getx}~r_1~n(r_2) &
\{ 
 r_1{{=}}\type{c}^0
 {;}\,
 r_2{=}\type{u}^{f}!n!\typevar{X}
 \}
       &// \small Load word from string\\
 \{r{=}\type{c}^{n^f}!\typevar{X}
      \}
&\opcode{stepx}~r~n &
\{ r{=}\type{c}^f!\typevar{Y}\}
       &// \small Step along string\\
 \{~ \}
&\opcode{newh}~r~n &
\{ r\,\,{=}\type{u}^n!\typevar{X}
      \}
       &// \small Set reg. $r$  content\\
 \{
 r_1{=}\type{c}^{f_1}!\typevar{Y}
 {;}\,
 r_2{=}\type{u}^{f_2}!\typevar{X}
      \}
&\opcode{swth}~r_1~n(r_2) &
\{
  r_1{=}\type{c}^{f_1}!\typevar{Y}
  {;}\,
  r_2{=}\type{u}^{f_2}!n!\typevar{X}
  \}
       &// \small Store word to array\\
 \{
 r_2{=}\type{u}^f!n!\typevar{X}
      \}
&\opcode{lwfh}~r_1~n(r_2) &
\{ 
 r_1{{=}}\type{c}^0
 {;}\,
 r_2{=}\type{u}^f!n!\typevar{X}\}
       &// \small Load word from array\\
 \{r_1{=}\typevar{x}
 {;}\,
 r_2^*{=}\type{c}^f!\typevar{X}
      \}
&\opcode{put}~r_1~n &
\{
 r_1{,}(n){=}\typevar{x}
 {;}\,
 r_2^*{=}\type{c}^f!n!\typevar{X}
 \}
       &// \small Store word to stack\\
 \{
 r_2^*{=}\type{c}^f!n!\typevar{X}
 {;}\,
 (n){=}\typevar{x}
 \}
&\opcode{get}~r_1~n &
\{
 r_1{,}(n){=}\typevar{x}
 {;}\,
 r_2^*{=}\type{c}^f!n!\typevar{X}
 \}
       &// \small Load word from stack\\
 \{r^*{=}\type{c}^{f}!\typevar{X}
      \}
&\opcode{push}~n &
\{ r^*{=}\type{c}^{n^f}\}
       &// \small New frame\\
 \{r_2^*{=}\type{c}^f!\typevar{X}
      \}
&\opcode{cspt}~r_1 &
\{ r_1{,}r_2^*{=}\type{c}^f!\typevar{X}\}
       &// \small Copy SP to reg. $r_1$\\
 \{r_1^*{=}\type{c}^f!\typevar{Y}{;\,}r_2{=}\type{c}^f!\typevar{X}
      \}
&\opcode{cspf}~r_2 &
\{ r_1^*{,}r_2{=}\type{c}^f!\typevar{X}\}
       &// \small Copy SP from reg. $r_2$\\
 \{r_1^*{=}\type{c}^{n^f}!\typevar{Y}{;\,}r_2{=}\type{c}^f!\typevar{X}
      \}
&\opcode{rspf}~r_2 &
\{ r_1^*{,}r_2{=}\type{c}^f!\typevar{X}\}
       &// \small Restore SP from reg. $r_2$\\
 \{~ \}
&\opcode{nop}           &
\{ ~\}
       &// \small No-op{,} do nothing\\
 \{r_2{=}\typevar{x}
      \}
&\opcode{mov}~r_1~r_2 &
\{ r_1{,}r_2{=}\typevar{x}\}
       &// \small Copy from reg. $r_2$\\
 \{r_2{=}\type{c}^f!\typevar{X}
      \}
&\opcode{addaiu}~r_1~r_2~n &
\{ 
    r_1{=}\type{c}^0
    {;}\,
    r_2{=}\type{c}^f!\typevar{X}
 \}
       &// \small Arithmetic add \\
\end{array}
\]
\begin{scriptsize}
\begin{notation}
The \typevar{X}, \typevar{Y}, etc 
stand for a set of offsets $!n_1!n_2!\dots$, for literal
natural numbers $n$. The stack frame size (or `tower of stack frame sizes')
$f$ is a literal natural number (or finite sequence of natural numbers).
The \typevar{x}, \typevar{y}, etc 
stand for any type (something that can appear on the right of an
equals sign).
\end{notation}
\end{scriptsize}
\end{table}

Small-step annotations $\{ \Theta \}~\kappa~\{ \Psi \}$ for an instruction
$\iota$ at address $a$ with a disassembly $\kappa$ generate a so-called `big step' rule
\[
\frac{
T~\triangleright~\{\Psi\}~a+4~\{\Phi\}
}{
T~\triangleright~\{\Theta\} ~a~ \{\Phi\}
}[a~|~\iota ~/~ \kappa]
\]
in which $\Phi$ is the final annotation at program end and $T$ denotes
a list of big-step annotations $\{ \Psi\}~a~\{\Phi\}$, one for each
instruction address $a$ in the program (note that, in consequence,
branches within the program must get the same annotation at
convergence as there is only one annotation there).  Thus the big-step
rule is an inference about what \emph{theory} $T$ contains.  The rule
above says that if $\{\Psi\}~a+4~\{\Phi\}$ is in theory $T$, then so is
$\{\Theta\} ~a~ \{\Phi\}$.  The label justifies the inference by the
fact that instruction $\iota$ is at address $a$, and disassembly
$\kappa$ has been chosen for it.

%
%

The big-step rules aim to generate a `covering' theory $T$ for each program.
That is, an annotation before every (reachable) instruction, and thus an
annotation {\em between} every instruction.  The rule above tells one
how to extend by one further instruction a theory that is growing from
the back of the program towards the front.

Where does theory construction start? It is with
the big-step rule for the final \opcode{jr}~\reg{ra} instruction
that classically ends a subroutine.
The action of this instruction is to jump back to the `return address' stored in
the \reg{ra} register (or another designated register).
The annotation for it says that there was a
program address (an `uncalculatable value', $\type{u}^0$) in the
\reg{ra} register before it ran (and it is still there after), and
requires no hypotheses:
$$
\frac{
}{
T~\triangleright~
\{r{=}\type{u}^0\} ~a~ \{r{=}\type{u}^0\}
}\mbox{[$a~|$~ \opcode{jr} $r$ / \opcode{return}]}
$$
The `$0$' superscript indicates that the address may not be used
as a base for offset memory accesses;  that would access program
instructions if it were allowed.
%
%
Calling code conventionally places the return address in the
\reg{ra} register prior to each subroutine call.


There are just three more big-step rules, corresponding to each
of the instructions that cause changes in the flow of control in
a program.  Jumps (unconditional branches) are handled by a rule
that refers back to the target of the jump:
\begin{align*}
\frac{
T ~\triangleright~ \{\Theta\} ~b~ \{\Phi\}
}{
T ~\triangleright~ \{\Theta\} ~a~ \{\Phi\}
}\mbox{[$a$ $|$ \opcode{j} $b$ / \opcode{goto} $b$]}
\end{align*}
This rule propagates the annotation at the
target $b$ of the jump back to the source $a$.
At worst a guess at the fixpoint is needed.

The logic of branch instructions (conditional jumps) at $a$ says that the
outcome of going down a branch to $b$ or continuing at $a+4$
must be the same. But the instruction
\opcode{bnez}~$r$~$b$ (`branch to address $b$ if register
$r$ is nonzero,
else continue') and variants first require the value in the
register $r$ to be tested, so it is pre-marked with \type{c}
(`calculatable'):
\begin{align*}
\frac{
T~\triangleright~ \{r{=}\type{c}^f!\typevar{X};\Theta\} ~b~ \{\Phi\}
\quad
T ~\triangleright~ \{r{=}\type{c}^f!\typevar{X};\Theta\} ~a+4~ \{\Phi\}
}{
T~\triangleright~ \{r{=}\type{c}^f!\typevar{X};\Theta\} ~a~ \{\Phi\}
}\mbox{[$a$ $|$ \opcode{bnez} $r$ $b$ / \opcode{ifnz} $r$ $b$]}
\end{align*}
The case $b< a$ (backward branch) requires a guess at a fixpoint
as it does for jump.
The annotated incremental history $f$, likely none, of
the value in the tested register is irrelevant here, but it is
maintained through the rule.  The set of offsets \typevar{X}
already written to is
also irrelevant here, but it is maintained through the rule. 


The RISC \opcode{jal}~$b$ machine code instruction implements
standard imperative programming language subroutine calls.  It
puts the address of the next instruction in the \reg{ra}
register (the `return address') and jumps to the subroutine at
address $b$.  The calling code will have saved the current
return address on the stack before the call.  The callee code
will return to the caller by jumping to the address in the
\reg{ra} register with \opcode{jr}~\reg{ra}, and the calling
code will then restore its own return address from the stack.

Because of  \opcode{jal}'s action in filling register \reg{ra}
with a program address, \reg{ra} on entry to the subroutine at
$b$ must already have a $\type{u}^0$ annotation, indicating an
unmodifiable value that cannot even be used for memory access.
And because the same subroutine can be called from many
different contexts, we need to distinguish the annotations per
call site and so we use a throwaway lettering $T'$ to denote
those annotations that derive from the call of $b$ from site
$a$.
The general rule is:
\begin{align*}
\frac{
T'~\triangleright~ \{\reg{ra}{=}\type{u}^0; \Psi\} ~b~ \{\Theta\}
\qquad
T ~\triangleright~ \{\Theta\} ~a+4~ \{\Phi\}
}{
T~\triangleright~ \{\Psi\} ~a~ \{\Phi\}
}\mbox{[$a$ $|$ \opcode{jal} $b$ / \opcode{gosub} $b$]}
\end{align*}
%
The `$0$' superscript means that memory accesses via the
return address as base address for \opcode{lw}/\opcode{sw}
are not allowed;  that would access the program instructions.
The stack pointer register has not been named, but
it must be distinct from the \reg{ra} register.

We have found it useful to apply extra constraints
at subroutine calls. We require (i) that each subroutine return
the stack to the same state it acquired it in (this is not a
universal convention), and (ii) that a subroutine make and
unmake all of its own local stack frame (again, not a universal
convention). That helps a Prolog implementation of the
verification logic start from a definitely known state at the
end of each subroutine independent of the call context --
namely, that the local stack frame at subroutine end (and beginning)
is size zero.
%
These constraints may be built into the \opcode{jal} rule as follows:
\begin{align*}
\frac{
T'~\triangleright~ \{\reg{ra} {=} \type{u}^0;
r^*{=}\type{c}^{0}!\typevar{X},\Psi\}
                                         ~b~
                                         \{r^*{=}\type{c}^{0}!\typevar{Y}; \Theta\}
\quad
T ~\triangleright~ \{r^*{=}\type{c}^{f}!\typevar{Y}; \Theta\}
                                         ~a{+}4~ \{\Phi\}
}{
T ~\triangleright~ \{r^*{=}\type{c}^{f}!\typevar{X}; \Psi\} ~a~ \{\Phi\}
}
\end{align*}
The requirement (i) is implemented by 
returning the stack pointer in the same register 
($r^*$ with the same $r$ on entry and return) and with no stack cells
visible in the local stack frame handed to the
subroutine and handed back by the subroutine (the two $0$s). The
requirement (ii) is implemented by setting the local stack frame on entry to
contain no stack, just the general purpose registers, which forces the
subroutine to make its own stack frame to work in.
%
%
Other calling conventions require other rule refinements.

As noted, the small-step and big-step rules can be read as a Prolog
program with variables the bold-faced offsets variables $\typevar{X}$,
$\typevar{Y}$, etc, and type variables $\typevar{x}$, $\typevar{y}$,
etc.

\section{Example annotation}
\label{sec:Is any substantive program allowed?}


Below is the annotation of the simple main routine of a Hello
World program that calls `printstr' with the Hello World string
address as argument, then calls `halt'.  The code was emitted by
a standard compiler ({\em gcc}) and modified by hand to be safe
against aliasing, so some compiler `quirks' are still visible.
The compiler likes to preserve the \reg{fp} register content
across subroutine calls, for example, even though it is not used
here.

The functionality is not at issue here, but, certainly, knowing
what each instruction does allows the annotation to be inferred
by an annotator without reference to rules and axioms.  The
\opcode{li}~\reg{a0} instruction sets the \reg{a0} (`$0$th
argument') register, for example, so the only change in the
annotation after the instruction is to the \reg{a0} column.  The
annotator introduces the string type, $\type{c}^{\ddot{1}}$,
into the annotation there, since the instruction sets \reg{a0}
to the address of the Hello World string.  The annotator assumes
that the stack pointer starts in the \reg{sp} register and that
`main' is called (likely from a set-up routine) with a return
address in the \reg{ra} register.  Changes are marked in grey:

\begin{center}\scriptsize
\begin{longtable}{lll|@{~}c@{~}|@{~}c@{~}|@{~}c@{~}|@{~}c@{~}|@{~}c@{~}|@{~}c@{~}|@{~}c@{~}|@{~}c@{~}|@{~}c@{~}|@{~}c@{~}}
  &
    &
      & \(\reg{sp}\sp*\)
        & \reg{ra}
          & \reg{a0}
            & \reg{fp}
              & \reg{gp}
                & \reg{v0}
                  & \reg{v1}
                    & (16)
                      & (24)
                        & (28)
\\
\hline
&&&&&&&&&&&\\[-1ex]
\underline{main}:   &\tt & &
    \(\type{c}\sp{0}\)&
    \(\type{u}\sp{0}\)&
    &
    \typevar{x}&
    &
    \(\type{c}\sp{\ddot{1}}!0\)&
    \(\type{c}\sp{0}\)&
    &
    \\
\tt move gp sp      &\rm cspt gp      &&
    \(\type{c}\sp{0}\)&
    \(\type{u}\sp{0}\)&
    &
    \typevar{x}&
    \cellcolor[gray]{0.8}\(\type{c}\sp{0}\)&
    \(\type{c}\sp{\ddot{1}}!0\)&
    \(\type{c}\sp{0}\)&
    &&
    \\
\tt addiu sp sp -32 &\rm push 32      &&
    \cellcolor[gray]{0.8} \(\type{c}\sp{32\sp0}\)&
    \(\type{u}\sp{0}\)&
    &
    \typevar{x}&
    \(\type{c}\sp{0}\)&
    \(\type{c}\sp{\ddot{1}}!0\)&
    \(\type{c}\sp{0}\)&
    &&
    \\
\tt sw ra 28(sp)    &\rm put  ra 28   &&
    \(\type{c}\sp{32\sp0} \cellcolor[gray]{0.8} !28\)&
    \(\type{u}\sp{0}\)&
    &
    \typevar{x}&
    \(\type{c}\sp{0}\)&
    \(\type{c}\sp{\ddot{1}}!0\)&
    \(\type{c}\sp{0}\)&
    &&
    \cellcolor[gray]{0.8} \(\type{u}\sp{0}\)
    \\
\tt sw fp 24(sp)    &\rm put  fp 24   &&
    \(\type{c}\sp{32\sp0} \cellcolor[gray]{0.8} !24!28\)&
    \(\type{u}\sp{0}\)&
    &
    \typevar{x}&
    \(\type{c}\sp{0}\)&
    \(\type{c}\sp{\ddot{1}}!0\)&
    \(\type{c}\sp{0}\)&
    &
    \cellcolor[gray]{0.8} \typevar{x}&
    \(\type{u}\sp{0}\)
    \\
\tt move fp sp      &\rm cspt fp      &&
    \(\type{c}\sp{32\sp0}!24!28\)&
    \(\type{u}\sp{0}\)&
    &
    \cellcolor[gray]{0.8}\(\type{c}\sp{32\sp0}!24!28\)&
    \(\type{c}\sp{0}\)&
    \(\type{c}\sp{\ddot{1}}!0\)&
    \(\type{c}\sp{0}\)&
    &
    \typevar{x}&
    \(\type{u}\sp{0}\)
    \\
\tt sw gp 16(sp)    &\rm put  gp 16   &&
    \(\type{c}\sp{32\sp0}!16!24!28\)&
    \(\type{u}\sp{0}\)&
    &
    \(\type{c}\sp{32\sp0}!24!28\)&
    \(\type{c}\sp{0}\)&
    \(\type{c}\sp{\ddot{1}}!0\)&
    \(\type{c}\sp{0}\)&
    \cellcolor[gray]{0.8}\(\type{c}\sp{0}\)&
    \typevar{x}&
    \(\type{u}\sp{0}\)
    \\
\tt li a0 <{\rm helloworld}>&\rm newx a0 \dots 1&&
    \(\type{c}\sp{32\sp0}!16!24!28\)&
    \(\type{u}\sp{0}\)&
    \cellcolor[gray]{0.8}\(\type{c}\sp{\ddot{1}}\) &
    \(\type{c}\sp{32\sp0}!24!28\)&
    \(\type{c}\sp{0}\)&
    \(\type{c}\sp{\ddot{1}}!0\)&
    \(\type{c}\sp{0}\)&
    \(\type{c}\sp{0}\)&
    \typevar{x}&
    \(\type{u}\sp{0}\)
    \\
\tt jal <{\rm printstr}>    &\rm gosub \dots  &&
    \(\type{c}\sp{32\sp0}!16!24!28\)&
    \cellcolor[gray]{0.8}\(\type{u}\sp{0}\)&
    \cellcolor[gray]{0.8}\(\type{c}\sp{0}\)&
    \(\type{c}\sp{32\sp0}!24!28\)&
    \cellcolor[gray]{0.8}\(\type{c}\sp{0}\)&
    \cellcolor[gray]{0.8}\(\type{c}\sp{0}\)&
    \cellcolor[gray]{0.8}\(\type{u}\sp{1}!0\)&
    \(\type{c}\sp{0}\)&
    \typevar{x}&
    \(\type{u}\sp{0}\)
    \\
\tt lw gp 16(sp)    &\rm get  gp 16   &&
    \(\type{c}\sp{32\sp0}!16!24!28\)&
    \(\type{u}\sp{0}\)&
    \(\type{c}\sp{0}\)&
    \(\type{c}\sp{32\sp0}!24!28\)&
    \cellcolor[gray]{0.8}\(\type{c}\sp{0}\)&
    \(\type{c}\sp{0}\)&
    \(\type{u}\sp{1}!0\)&
    \(\type{c}\sp{0}\)&
    \typevar{x}&
    \(\type{u}\sp{0}\)
    \\
\tt jal <{\rm halt}>      &\rm gosub \dots  &&
    \(\type{c}\sp{32\sp0}!16!24!28\)&
    \cellcolor[gray]{0.8}\(\type{u}\sp{0}\)&
    \(\type{c}\sp{0}\)&
    \(\type{c}\sp{32\sp0}!24!28\)&
    \(\type{c}\sp{0}\)&
    \(\type{c}\sp{0}\)&
    \cellcolor[gray]{0.8}\(\type{u}\sp{1}!0\)&
    \(\type{c}\sp{0}\)&
    \typevar{x}&
    \(\type{u}\sp{0}\)
    \\
\tt nop             &\rm              &  &&&&&&&&&\\
\tt lw gp 16(sp)    &\rm get  gp 16   &&
    \(\type{c}\sp{32\sp0}!16!24!28\)&
    \(\type{u}\sp{0}\)&
    \(\type{c}\sp{0}\)&
    \(\type{c}\sp{32\sp0}!24!28\)&
    \cellcolor[gray]{0.8}\(\type{c}\sp{0}\)&
    \(\type{c}\sp{0}\)&
    \(\type{u}\sp{1}!0\)&
    \(\type{c}\sp{0}\)&
    \typevar{x}&
    \(\type{u}\sp{0}\)
    \\
\tt nop             &\rm              &  &&&&&&&&&\\
\tt lw ra 28(sp)    &\rm get  ra 28   &&
    \(\type{c}\sp{32\sp0}!16!24!28\)&
    \cellcolor[gray]{0.8}\(\type{u}\sp{0}\)&
    \(\type{c}\sp{0}\)&
    \(\type{c}\sp{32\sp0}!24!28\)&
    \(\type{c}\sp{0}\)&
    \(\type{c}\sp{0}\)&
    \(\type{u}\sp{1}!0\)&
    \(\type{c}\sp{0}\)&
    \typevar{x}&
    \(\type{u}\sp{0}\)
    \\
\tt lw fp 24(sp)    &\rm get  fp 24   &&
    \(\type{c}\sp{32\sp0}!16!24!28\)&
    \(\type{u}\sp{0}\)&\(\type{c}\sp{0}\)&
    \cellcolor[gray]{0.8}\(\typevar{x}\)&
    \(\type{c}\sp{0}\)&
    \(\type{c}\sp{0}\)&
    \(\type{u}\sp{1}!0\)&
    \(\type{c}\sp{0}\)&
    \typevar{x}&
    \(\type{u}\sp{0}\)
    \\
\tt move sp gp      &\rm rspf gp      &&
    \cellcolor[gray]{0.8}\(\type{c}\sp{0}\)&
    \(\type{u}\sp{0}\)&
    \(\type{c}\sp{0}\)&
    \(\typevar{x}\)&
    \(\type{c}\sp{0}\)&
    \(\type{c}\sp{0}\)&
    \(\type{u}\sp{1}!0\)&
    \(\type{c}\sp{0}\)&
    \typevar{x}&
    \(\type{u}\sp{0}\)
    \\
\tt jr ra           &\rm return       &&
    \(\type{c}\sp{0}\)&
    \cellcolor[gray]{0.8}\(\type{u}\sp{0}\)&
    \(\type{c}\sp{0}\)&
    \(\typevar{x}\)&
    \(\type{c}\sp{0}\)&
    \(\type{c}\sp{0}\)&
    \(\type{u}\sp{1}!0\)&
    \(\type{c}\sp{0}\)&
    \typevar{x}&
    \(\type{u}\sp{0}\)
    \\
\underline{helloworld}:     &\rm $\langle$string data$\rangle$ &&&&&&&&&&&\\[-4ex]
\end{longtable}
\end{center}
That the `!' annotations are always less than the bottom element of the 
tower on the stack pointer annotation
means that no aliasing occurs.
Reads are at an offset already marked
with a `!', hence within the same range that writes are constrained to. 

%
%

The `halt' subroutine does not use the stack pointer; its
function is to write a single  byte to the hard-coded I/O-mapped
address of a system peripheral.
%
%
The annotation for register \reg{v1} on output is the taint left by
that write.


%
%
\begin{scriptsize}
\begin{leftprogram}
\fns \underline{halt}:                             \#\rm \(\reg{zero}=\type{c}\sp0;\reg{ra}=\type{u}\sp0\)
\fns\tt li v1 0xb0000x10   newh v1 ... 1  \#\rm \(\reg{v1}=\type{u}\sp1;\reg{zero}=\type{c}\sp0;\reg{ra}=\type{u}\sp0\)
\fns\tt sb zero 0(v1)      sbth v1 0(v1)  \#\rm \(\reg{v1}=\type{u}\sp1!0;\reg{zero}=\type{c}\sp0;\reg{ra}=\type{u}\sp0\)
\fns\tt jr ra              return         \#\rm \(\reg{v1}=\type{u}\sp1!0;\reg{zero}=\type{c}\sp0;\reg{ra}=\type{u}\sp0\)
\end{leftprogram}
\end{scriptsize}

\noindent
The \reg{zero} register is conventionally kept filled with the zero word in
RISC architectures.

The \emph{printstr} routine takes a string pointer as argument in
register \reg{a0}.  A requirement that registers \reg{v0}, \reg{v1}
have certain types on entry is an artifact of annotation. 
Since `\$B' comes after writes to \reg{v0}, \reg{v1}, those two registers
are bound to types at that point.  The forward jump (\opcode{j}) to 
`\$B' forces the same annotations at the jump instruction as at the
target.  But, at the jump, no write to \reg{v0}, \reg{v1}
has yet taken place, so we are obliged to provide
the types of \reg{v0}, \reg{v1} at entry. The table below is
constructed using the same display convention as the table for {\it
main}.
%
\begin{center}
\scriptsize
\begin{longtable}{@{}lllc|c|c|c|c|c|c|c|c|c|c@{}}
    && &
    \reg{sp}$^*$&
    \reg{fp}&
    \reg{ra}&
    \reg{a0}&
    \reg{gp}&
    \reg{v0}&
    \reg{v1}&
    (12)&
    (20)&
    (24)&
    (28)
    \\
\hline
    &&& &&&&&&&&&
    \\[-1.5ex]
\underline{printstr}:&                   &\#&\rm
    \(\type{c}\sp0\)&
    \typevar{x}&
    \(\type{u}\sp0\)&
    \(\type{c}\sp{\ddot{1}}!0\)&
    &
    \(\type{c}\sp{\ddot{1}}!0\)&
    \(\type{u}\sp1!0\)&
    &&&
    \\
\tt move gp sp   &     cspt gp   &\#&\rm
    \(\type{c}\sp0\)&
    \typevar{x}&
    \(\type{u}\sp0\)&
    \(\type{c}\sp{\ddot{1}}!0\)&
    \cellcolor[gray]{0.8}\(\type{c}\sp0\)&
    \(\type{c}\sp{\ddot{1}}!0\)&
    \(\type{u}\sp1!0\)&
    &&&
    \\
\tt addiu sp sp -32 &  push 32   &\#&\rm
    \cellcolor[gray]{0.8}\(\type{c}\sp{32\sp0}\)&
    \typevar{x}&
    \(\type{u}\sp0\)&
    \(\type{c}\sp{\ddot{1}}!0\)&
    \(\type{c}\sp0\)&
    \(\type{c}\sp{\ddot{1}}!0\)&
    \(\type{u}\sp1!0\)&
    &&& 
    \\
\tt sw ra 24(sp)    &  put ra 24 &\#&\rm
    \cellcolor[gray]{0.8}\(\type{c}\sp{32\sp0}!24\)&
    \typevar{x}&
    \(\type{u}\sp0\)&
    \(\type{c}\sp{\ddot{1}}!0\)&
    \(\type{c}\sp0\)&
    \(\type{c}\sp{\ddot{1}}!0\)&
    \(\type{u}\sp1!0\)&
    &&  
    \cellcolor[gray]{0.8}\(\type{u}\sp0\)&
    \\
\tt sw fp 20(sp)    &  put fp 20 &\#&\rm
    \cellcolor[gray]{0.8}\(\type{c}\sp{32\sp0}!20!24\)&
    \typevar{x}&
    \(\type{u}\sp0\)&
    \(\type{c}\sp{\ddot{1}}!0\)&
    \(\type{c}\sp0\)&
    \(\type{c}\sp{\ddot{1}}!0\)&
    \(\type{u}\sp1!0\)&
    & 
    \cellcolor[gray]{0.8}\typevar{x}&
    \(\type{u}\sp0\)&
    \\
\tt move fp sp      &  cspt fp   &\#&\rm
    \(\type{c}\sp{32\sp0}!20!24\)&
    \cellcolor[gray]{0.8}\(\type{c}\sp{32\sp0}!20!24\)&
    \(\type{u}\sp0\)&
    \(\type{c}\sp{\ddot{1}}!0\)&
    \(\type{c}\sp0\)&
    \(\type{c}\sp{\ddot{1}}!0\)&
    \(\type{u}\sp1!0\)&
    & 
    \typevar{x}&
    \(\type{u}\sp0\)&
    \\
\tt sw gp 12(sp)    &  put gp 12 &\#&\rm
    \cellcolor[gray]{0.8}\(\type{c}\sp{32\sp0}!12!20!24\)&
    \(\type{c}\sp{32\sp0}!20!24\)&
    \(\type{u}\sp0\)&
    \(\type{c}\sp{\ddot{1}}!0\)&
    \(\type{c}\sp0\)&
    \(\type{c}\sp{\ddot{1}}!0\)&
    \(\type{u}\sp1!0\)&
    \cellcolor[gray]{0.8}\(\type{c}\sp0\)&
    \typevar{x}&
    \(\type{u}\sp0\)&
     \\
\tt sw a0 28(sp)    &  put a0 28 &\#&\rm
    \cellcolor[gray]{0.8}\(\type{c}\sp{32\sp0}!12!20!24!28\)&
    \(\type{c}\sp{32\sp0}!20!24\)&
    \(\type{u}\sp0\)&
    \(\type{c}\sp{\ddot{1}}!0\)&
    \(\type{c}\sp0\)&
    \(\type{c}\sp{\ddot{1}}!0\)&
    \(\type{u}\sp1!0\)&
    \(\type{c}\sp0\)&
    \typevar{x}&
    \(\type{u}\sp0\)&
    \cellcolor[gray]{0.8}\(\type{c}\sp{\ddot{1}}!0\)
     \\
\tt move a0 zero    &  mov a0 zero&\#&\rm
    \(\type{c}\sp{32\sp0}!12!20!24!28\)&
    \(\type{c}\sp{32\sp0}!20!24\)&
    \(\type{u}\sp0\)&
    \cellcolor[gray]{0.8}\(\type{c}\sp0\)&
    \(\type{c}\sp0\)&
    \(\type{c}\sp{\ddot{1}}!0\)&
    \(\type{u}\sp1!0\)&
    \(\type{c}\sp0\)&
    \typevar{x}&
    \(\type{u}\sp0\)&
    \(\type{c}\sp{\ddot{1}}!0\)
     \\
\tt j \(\rm\langle\$B\rangle\)    & j \(\rm\langle\$B\rangle\)      &\#&\rm
     &
     &
     &
     &
     &
&&&&
     \\
\underline{\$A}:    &             &\#&\rm
    \(\type{c}\sp{32\sp0}!12!20!24!28\)&
    \(\type{c}\sp{32\sp0}!20!24\)&
    \(\type{u}\sp0\)&
    \(\type{c}\sp0\)&
    \(\type{c}\sp0\)&
    \cellcolor[gray]{0.8}\(\type{c}\sp0\)&
    \(\type{u}\sp1!0\)&
    \(\type{c}\sp0\)&
    \typevar{x}&
    \(\type{u}\sp0\)&
    \(\type{c}\sp{\ddot{1}}!0\)
    \\
\tt lw v0 28(sp)   &   get v0 28 &\#&\rm
    \(\type{c}\sp{32\sp0}!12!20!24!28\)&
    \(\type{c}\sp{32\sp0}!20!24\)&
    \(\type{u}\sp0\)&
    \(\type{c}\sp0\)&
    \(\type{c}\sp0\)&
    \cellcolor[gray]{0.8}\(\type{c}\sp{\ddot{1}}!0\)&
    \(\type{u}\sp1!0\)&
    \(\type{c}\sp0\)&
    \typevar{x}&
    \(\type{u}\sp0\)&
    \(\type{c}\sp{\ddot{1}}!0\)
    \\
\tt nop            & nop         &\#&\rm
     &
     &
     &
     &
     &
     &&&&
     \\
\tt lb v0 0(v0)    &   getbx v0 0(v0) \kern-2.5pt&\#&\rm
    \(\type{c}\sp{32\sp0}!12!20!24!28\)&
    \(\type{c}\sp{32\sp0}!20!24\)&
    \(\type{u}\sp0\)&
    \(\type{c}\sp0\)&
    \(\type{c}\sp0\)&
    \cellcolor[gray]{0.8}\(\type{c}\sp0\)&
    \(\type{u}\sp1!0\)&
    \(\type{c}\sp0\)&
    \typevar{x}&
    \(\type{u}\sp0\)&
    \(\type{c}\sp{\ddot{1}}!0\)
    \\
\tt move v1 v0     &   mov v1 v0 &\#&\rm
    \(\type{c}\sp{32\sp0}!12!20!24!28\)&
    \(\type{c}\sp{32\sp0}!20!24\)&
    \(\type{u}\sp0\)&
    \(\type{c}\sp0\)&
    \(\type{c}\sp0\)&
    \(\type{c}\sp0\)&
    \cellcolor[gray]{0.8}\(\type{c}\sp0\)&
    \(\type{c}\sp0\)&
    \typevar{x}&
    \(\type{u}\sp0\)&
    \(\type{c}\sp{\ddot{1}}!0\)
    \\
\tt lw v0 28(sp)   &   get v0 28 &\#&\rm
    \(\type{c}\sp{32\sp0}!12!20!24!28\)&
    \(\type{c}\sp{32\sp0}!20!24\)&
    \(\type{u}\sp0\)&
    \(\type{c}\sp0\)&
    \(\type{c}\sp0\)&
    \cellcolor[gray]{0.8}\(\type{c}\sp{\ddot{1}}!0\)&
    \(\type{c}\sp0\)&
    \(\type{c}\sp0\)&
    \typevar{x}&
    \(\type{u}\sp0\)&
    \(\type{c}\sp{\ddot{1}}!0\)
    \\
\tt addiu v0 v0 1  &   step v0 1 & \#&\rm
    \(\type{c}\sp{32\sp0}!12!20!24!28\)&
    \(\type{c}\sp{32\sp0}!20!24\)&
    \(\type{u}\sp0\)&
    \(\type{c}\sp0\)&
    \(\type{c}\sp0\)&
    \cellcolor[gray]{0.8}\(\type{c}\sp{\ddot{1}}!0\)&
    \(\type{c}\sp0\)&
    \(\type{c}\sp0\)&
    \typevar{x}&
    \(\type{u}\sp0\)&
    \(\type{c}\sp{\ddot{1}}!0\)
    \\
\tt sw v0 28(sp)   &   put v0 28 &\#&\rm
    \(\type{c}\sp{32\sp0}!12!20!24!28\)&
    \(\type{c}\sp{32\sp0}!20!24\)&
    \(\type{u}\sp0\)&
    \(\type{c}\sp0\)&
    \(\type{c}\sp0\)&
    \cellcolor[gray]{0.8}\(\type{c}\sp{\ddot{1}}!0\)&
    \(\type{c}\sp0\)&
    \(\type{c}\sp0\)&
    \typevar{x}&
    \(\type{u}\sp0\)&
    \(\type{c}\sp{\ddot{1}}!0\)
    \\
\tt move a0 v1     &   mov a0 v1 &\#&\rm
    \(\type{c}\sp{32\sp0}!12!20!24!28\)&
    \(\type{c}\sp{32\sp0}!20!24\)&
    \(\type{u}\sp0\)&
    \cellcolor[gray]{0.8}\(\type{c}\sp0\)&
    \(\type{c}\sp0\)&
    \(\type{c}\sp{\ddot{1}}!0\)&
    \(\type{c}\sp0\)&
    \(\type{c}\sp0\)&
    \typevar{x}&
    \(\type{u}\sp0\)&
    \(\type{c}\sp{\ddot{1}}!0\)
    \\
\tt jal \(\rm\langle{\rm printchar}\rangle\)    &gosub printchar         & \#&\rm
    \(\type{c}\sp{32\sp0}!12!20!24!28\)&
    \(\type{c}\sp{32\sp0}!20!24\)&
    \cellcolor[gray]{0.8}\(\type{u}\sp0\)&
    \(\type{c}\sp0\)&
    \(\type{c}\sp0\)&
    \(\type{c}\sp{\ddot{1}}!0\)&
    \cellcolor[gray]{0.8}\(\type{u}\sp1!0\)&
    \(\type{c}\sp0\)&
    \typevar{x}&
    \(\type{u}\sp0\)&
    \(\type{c}\sp{\ddot{1}}!0\)
    \\
\tt lw gp 12(sp)   &   get gp 12 &\#&\rm
    \(\type{c}\sp{32\sp0}!12!20!24!28\)&
    \(\type{c}\sp{32\sp0}!20!24\)&
    \(\type{u}\sp0\)&
    \(\type{c}\sp0\)&
    \cellcolor[gray]{0.8}\(\type{c}\sp0\)&
    \(\type{c}\sp{\ddot{1}}!0\)&
    \(\type{u}\sp1!0\)&
    \(\type{c}\sp0\)&
    \typevar{x}&
    \(\type{u}\sp0\)&
    \(\type{c}\sp{\ddot{1}}!0\)
    \\
\underline{\$B}:   &             & \#&\rm
     &
     &
     &
     &
     &
&&&&
     \\
\tt lw v0 28(sp)   &   get v0 28 &\#&\rm
    \(\type{c}\sp{32\sp0}!12!20!24!28\)&
    \(\type{c}\sp{32\sp0}!20!24\)&
    \(\type{u}\sp0\)&
    \(\type{c}\sp0\)&
    \(\type{c}\sp0\)&
    \cellcolor[gray]{0.8}\(\type{c}\sp{\ddot{1}}!0\)&
    \(\type{u}\sp1!0\)&
    \(\type{c}\sp0\)&
    \typevar{x}&
    \(\type{u}\sp0\)&
    \(\type{c}\sp{\ddot{1}}!0\)
    \\
\tt lb v0 0(v0)    &   getbx v0 0(v0)&\#&\rm
    \(\type{c}\sp{32\sp0}!12!20!24!28\)&
    \(\type{c}\sp{32\sp0}!20!24\)&
    \(\type{u}\sp0\)&
    \(\type{c}\sp0\)&
    \(\type{c}\sp0\)&
    \cellcolor[gray]{0.8}\(\type{c}\sp0\)&
    \(\type{u}\sp1!0\)&
    \(\type{c}\sp0\)&
    \typevar{x}&
    \(\type{u}\sp0\)&
    \(\type{c}\sp{\ddot{1}}!0\)
    \\
\tt bnez v0 \(\rm\langle\$A\rangle\)&bnez v0 \(\rm\langle\$A\rangle\) &\#&\rm
    \(\type{c}\sp{32\sp0}!12!20!24!28\)&
    \(\type{c}\sp{32\sp0}!20!24\)&
    \(\type{u}\sp0\)&
    \(\type{c}\sp0\)&
    \(\type{c}\sp0\)&
    \(\type{c}\sp0\)&
    \(\type{u}\sp1!0\)&
    \(\type{c}\sp0\)&
    \typevar{x}&
    \(\type{u}\sp0\)&
    \(\type{c}\sp{\ddot{1}}!0\)
     \\
\tt move sp fp     &   cspf fp   &\#&\rm
    \cellcolor[gray]{0.8}\(\type{c}\sp{32\sp0}!20!24\)&
    \(\type{c}\sp{32\sp0}!20!24\)&
    \(\type{u}\sp0\)&
    \(\type{c}\sp0\)&
    \(\type{c}\sp0\)&
    \(\type{c}\sp0\)&
    \(\type{u}\sp1!0\)&
    \(\type{c}\sp0\)&
    \typevar{x}&
    \(\type{u}\sp0\)&
    \(\type{c}\sp{\ddot{1}}!0\)
     \\
\tt lw ra 24(sp)   &   get ra 24 &\#&\rm
    \(\type{c}\sp{32\sp0}!20!24\)&
    \(\type{c}\sp{32\sp0}!20!24\)&
    \cellcolor[gray]{0.8}\(\type{u}\sp0\)&
    \(\type{c}\sp0\)&
    \(\type{c}\sp0\)&
    \(\type{c}\sp0\)&
    \(\type{u}\sp1!0\)&
    \(\type{c}\sp0\)&
    \typevar{x}&
    \(\type{u}\sp0\)&
    \(\type{c}\sp{\ddot{1}}!0\)
     \\
\tt lw fp 20(sp)   &   get fp 20 &\#&\rm
    \(\type{c}\sp{32\sp0}!20!24\)&
    \cellcolor[gray]{0.8}\typevar{x}&
    \(\type{u}\sp0\)&
    \(\type{c}\sp0\)&
    \(\type{c}\sp0\)&
    \(\type{c}\sp0\)&
    \(\type{u}\sp1!0\)&
    \(\type{c}\sp0\)&
    \typevar{x}&
    \(\type{u}\sp0\)&
    \(\type{c}\sp{\ddot{1}}!0\)
     \\
\tt move sp gp     &   rspf gp   &\#&\rm
    \cellcolor[gray]{0.8}\(\type{c}\sp0\)&
    \typevar{x}&
    \(\type{u}\sp0\)&
    \(\type{c}\sp0\)&
    \(\type{c}\sp0\)&
    \(\type{c}\sp0\)&
    \(\type{u}\sp1!0\)&
    \(\type{c}\sp0\)&
    \typevar{x}&
    \(\type{u}\sp0\)&
    \(\type{c}\sp{\ddot{1}}!0\)
     \\
\tt jr ra          & return      &\#&\rm
    \(\type{c}\sp0\)&
    \typevar{x}&
    \(\type{u}\sp0\)&
    \(\type{c}\sp0\)&
    \(\type{c}\sp0\)&
    \(\type{c}\sp0\)&
    \(\type{u}\sp1!0\)&
    \(\type{c}\sp0\)&
    \typevar{x}&
    \(\type{u}\sp0\)&
    \(\type{c}\sp{\ddot{1}}!0\)
\end{longtable}
\end{center}

\noindent
The `printchar' subroutine writes a character received in
register \reg{a0} to the hard-coded address of a printer device:

\begin{scriptsize}
\begin{leftprogram}
\fns \underline{printchar}:                        \#\rm \(\reg{a0}=\type{c}\sp0;\reg{ra}=\type{u}\sp0\)
\fns\tt li v1 0xb0000000   newh v1 ... 1  \#\rm \(\reg{v1}=\type{u}\sp1;\reg{a0}=\type{c}\sp0;\reg{ra}=\type{u}\sp0\)
\fns\tt sb a0 0(v1)        sbth a0 0(v1)  \#\rm \(\reg{v1}=\type{u}\sp1!0;\reg{a0}=\type{c}\sp0;\reg{ra}=\type{u}\sp0\)
\fns\tt jr ra              return         \#\rm \(\reg{v1}=\type{u}\sp1!0;\reg{a0}=\type{c}\sp0;\reg{ra}=\type{u}\sp0\)
\end{leftprogram}
\end{scriptsize}

\noindent
Like \emph{halt}, it does not use the stack pointer.

\section{How does annotation ensure  aliasing  does not happen?}
\label{sec:Preventing unsafe memory access}

How to ensure
memory aliasing does not happen is intuitively simple: 
make sure that each address used
can have been calculated in only one way.  There are in principle two 
constraints that can be enforced directly via annotation and which
will have this effect:
\begin{enumerate}[(i)]
\item Both stack reads and writes with \opcode{get} and \opcode{put} may be
restricted to offsets $n$ that lie in the range permitted by the
local stack frame size (look for a stack pointer tower $m^{\nedots}$ on the
annotation before the instruction, with $0 \le n\le m-4$);
\item stack reads with  \opcode{get} may be restricted to offsets $n$ at
which writes with \opcode{put} have already taken place (look for a
$!n$ mark on the annotation before the instruction).
\end{enumerate}
Similarly for strings and arrays.
It is (i) that makes memory aliasing
impossible, but (ii) is also useful because it (a) reduces 
(i) to be required on writes alone, and (b) prevents `read
before write' faults. Without (i), code could validly try to access an
element of the caller's frame, and that would fail because of
aliasing via two distinct calculations for the same address, 
from caller's and callee's frames respectively.


If these constraints  are satisfied, we argue as follows that memory-aliasing 
cannot occur. The base address used for access via the RISC \opcode{lw} or
\opcode{sw} instructions is either:
\begin{enumerate}
\item The stack pointer (disassembly of the access instruction is to
\opcode{put}, \opcode{get}, \opcode{putb}, \opcode{getb});
\item the base address of a string, incremented several times by the
string increment (the disassembly is to
\opcode{putx}, \opcode{getx}, \opcode{putbx}, \opcode{getbx});
\item the base address of an array (the disassembly is to
\opcode{swth}, \opcode{sbth}, \opcode{lwfh}, \opcode{lbfh}).
\end{enumerate}
and the offset in the instruction is in the first case less than the
stack frame size, in the second case less than the string increment, and
in the third case less than the array size.

Why are these and no other case possible?  Firstly, if the program is
annotated, then every use of a base address for the underlying machine
code \opcode{lw} and \opcode{sw} instructions matches exactly one of
these cases, because the annotation rules have no other option.

Next we claim that the annotations on a program are {\em sound}.  This
is a technical claim that we cannot formally substantiate here that says
that in an annotated program the annotations around each instruction
reflect what the instruction does computationally.  The full statement
requires a model of each instruction's semantics as a state-to-state
transformation (given in Appendix~\ref{Sec:Motivating semantics}) and a
proof that the big-step rules of Section~\ref{sec:Formal logic} express
those semantics.
Given that, the three cases above for the base address used in a
\opcode{lw} and \opcode{sw} instruction may be characterized thus:
\begin{enumerate}
\item It is the stack pointer, which is marked with an asterisk in the
annotatiion and typed with
$\type{c}^f$ where the tower $f$ consists of  the sizes of current and
calling stack frames;
\item it is a string pointer, which  is typed with $\type{c}^{\ddot{m}}$
in the annotation and is equal to the base address of the string plus a
finite number of increments $m$;
\item it is an array pointer, which is typed with $\type{u}^m$ in the
annotation and is equal to the base address of the array, which is of
size $m$.
\end{enumerate}

\noindent
In each of those three cases, the offset used in the \opcode{lw} or
\opcode{sw} instruction is only permitted by the annotation to lie in
the range $0$ to $m-4$, where $m$ is respectively the current frame
size, the string step size, and the array size. The first of these
cases implements condition (i), and the second and third implement the
equivalent condition for strings and arrays respectively. I.e., there is only
one calculation possible for each address used.


Similar arguments hold for byte-wise access via \opcode{lb}
and \opcode{sb}. In addition, however, one must require that menory areas
accessed via these instructions are not also accessed via \opcode{lw}
and \opcode{sw}, in order to avoid different calculations for the
addresses of the individual bytes in a word.  The simplest way to ensure
that is to forbid use of \opcode{lb} and \opcode{sb} entirely, relying
instead on \opcode{lw} and \opcode{sw} plus arithmetic operations to
extract the byte.  The next simplest alternative is to allow \opcode{lb}
and \opcode{sb} only on strings with step size less than $4$ and arrays
of size less than 4, which word-wise instructions are forbidden from
accessing by the annotation rules.


\section{Conclusion and Future Work}

We have set out a method of annotation that can ensure that a RISC
machine-code program is safe against `hardware' aliasing.  We model
aliasing as introduced by the use of different arithmetic calculations
for the same memory address, and successful annotation guarantees that a
unique calculation will be used at run-time for the address of each
execution stack, string or array element accessed by the program.
Annotation also means disassembling the machine code to a slightly
higher level assembly language, for a stack machine, and a human being
is required to certify that the disassembly matches the programmer's
intentions. 

Note that one may add disassembly rules to the system that are
(deliberately) semantically wrong, with the aim of correcting the
code.  For example, one may choose to (incorrectly) disassemble the RISC
$\opcode{addiu}~\reg{sp}~\reg{sp}~32$ instruction to a stack machine
\opcode{pop} instruction.  The RISC instruction is not a correct
implementation of the higher level instruction in an aliasing context,
although it was likely intended to be.  But one may then replace the
original RISC code with a correct implementation.

Also note that the equational annotations here may be
generalised to quite arbitrary first-order predicates.
%
It also appears that our system of types may be generalised to arrays of
arrays and strings of strings, etc, which offers the prospect of
a static analysis technology that can follow pointers.



\bibliographystyle{plain}
\bibliography{asmtypes-OC}

\newpage
\appendix

\section{Motivating semantics}
\label{Sec:Motivating semantics}


We will restrict the commentary here to the ten instructions
from the 32-bit RISC instruction set architecture shown in
Table~\ref{Tab:3}.  These are also the elements of a tiny
RISC-16 machine code/assembly language \mycite{RISC16}.  Because
of their role in RISC-16, we know that they form a complete set
that can perform arbitrary computations.

%
We suppose in this paper that programs are such that the stack pointer
always remains in the \reg{sp} register.  Copies may be made of it
elsewhere using the \opcode{move} (copy) instruction, and it may be
altered in situ using the \opcode{addiu} instruction.  Adding a negative
amount increases the stack size, and stack conventionally grows
top-down in the address space. We also suppose that the {\em return
address} pointer is always in the source register $r$ at the point where
a (`jump register') $\opcode{jr}~r$ instruction is executed, so that the
latter may be interpreted as a stack machine \opcode{return}
instruction.

A program induces a set of {\em dataflow traces} through registers.
A dataflow trace is a unique path through registers and stack
memory cells that traces movement of data.
The segments of the trace may be labelled with {\em events}
as detailed below, signifying data transformation,  or they may be
unlabelled, signifying transfer without transformation.
Each trace starts with the introduction of a value into a register,
either from the instruction itself in the case of the \opcode{li}
and the source is shown as a blank
triangle, or by hypothesis at the start of a subroutine and the source
is shown as a vertical bar.
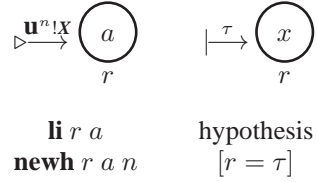
\[
\begin{array}[b]{ccc}
\triangleright\kern-4pt&
\mathop{\longrightarrow}\limits^{\mbox{\type{u}}^n!\typevar{X}} &\hbox
to 2em{\hss\hbox to 0pt{\hss\Huge$\bigcirc$\hss}\hbox to 0pt{\hss
\raisebox{3pt}{$a$}\hss}\hss}\\
& & r\\[2ex]
\multicolumn{3}{c}{\mbox{\rm \opcode{li}~$r$~$a$}}\\
\multicolumn{3}{c}{\mbox{\rm \opcode{newh}~$r$~$a$~$n$}}
\end{array}
\qquad
\begin{array}[b]{ccc}
{\vert}\kern-6pt&
\mathop{\longrightarrow}\limits^{\tau} &\hbox to 2em{\hss\hbox to
0pt{\hss\Huge$\bigcirc$\hss}\hbox to 0pt{\hss \raisebox{3pt}{$x$}\hss}\hss}\\
& & r\\[2ex]
\multicolumn{3}{c}{\mbox{hypothesis}}\\
\multicolumn{3}{c}{[r=\tau]}
\end{array}
\]
The left hand diagram above shows the introduction of the
address $a$ of an array of size $n$ into register $r$, the \opcode{li}
machine code instruction having been disassembled to \opcode{newh}.
The indices of those elements already written to the array are recorded
in the set $\typevar{X}$.  Usually that is the full set of indices up
to $n$ and the address is that of an array written earlier.
The label on the arrow is a {\em annotated type} (Table~ \ref{tab:ann}),
indicating an introduction event. The annotated type brought in with the array
pointer introduction is
\[
\type{u}^n!\typevar{X}
\]
standing for an address that may not subsequently be altered
(`\type{u}', or `uncalculatable') of $n$ bytes of memory, that has been
written to at each of the offsets in the set \typevar{X}.

If the \opcode{li} instruction is instead interpreted as introducing
the address $a$ of a `string-like' object, then the annotated type brought in is
\[
\type{c}^{\ddot{n}}!\typevar{X}
\]
standing for an address that may be altered (`\type{c}', or
`calculatable') and stepped in increments of $n$ bytes. The \typevar{X}
again stands for a set of offsets from the base (up to $n$ bytes)
at which the structure has been written. The same pattern \typevar{X}
applies at every increment $n$ along the string. The form `$\ddot{n}$'
is meant to be understood as `$n^{n^\nedots}$\kern3pt', with the $n$
repeated an indefinite number of times. This may be viewed as a variant of the
annotated type
\[
\type{c}^{n_1^{{\nedots}^{\kern3pt\raisebox{4pt}{\tiny$n_k$}}}}!\typevar{X}
\]
that the stack pointer is associated with (for some finite sequence
$n_1$,\dots,$n_k$ as superscripts) and which records a historical
sequence of local stack frames created one within the scope of the
other culminating in a current stack frame of size $n_1$ bytes.

\begin{figure}[tb]
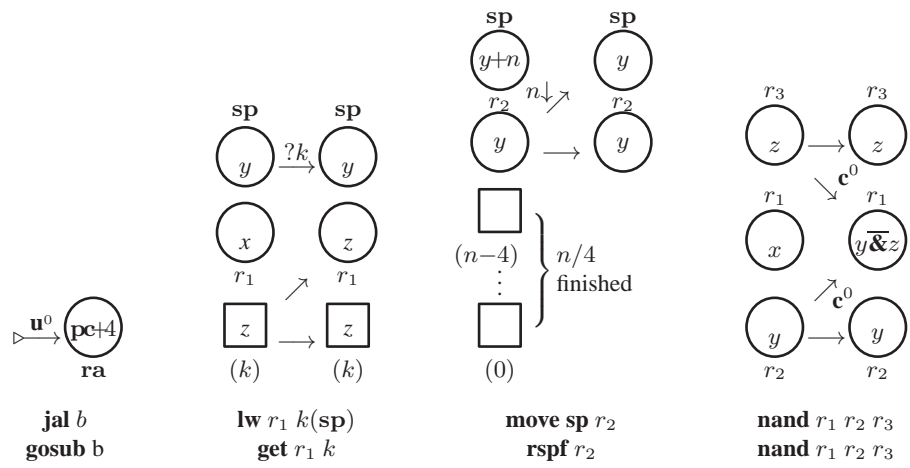

\[
\begin{array}{c@{\qquad\qquad}c@{\qquad\qquad}c@{\qquad\qquad}c}
\begin{array}[b]{ccc}
\triangleright\kern-4pt& \mathop{\longrightarrow}\limits^{\mbox{\type{u}}^0} &\hbox to
2em{\hss\hbox to 0pt{\hss\Huge$\bigcirc$\hss}\hbox to 0pt{\hss
\raisebox{3pt}{$\reg{p\kern-1pt c}{\kern-2pt+\kern-2pt}4$}\hss}\hss}\\
& & \reg{ra}\\[2ex]
\multicolumn{3}{c}{\mbox{\rm \opcode{jal}~$b$}}\\
\multicolumn{3}{c}{\mbox{\rm \opcode{gosub}~b}}
\end{array}
&
\begin{array}[b]{ccc}
{\reg{sp}} &   & {\reg{sp}}\\[0.2ex]
\hbox to 2em{\hss\hbox to 0pt{\hss\Huge$\bigcirc$\hss}\hbox to 0pt{\hss
$y$\hss}\hss}& \mathop{\longrightarrow}\limits^{\mbox{$?k$}} &\hbox to 2em{\hss\hbox to 0pt{\hss\Huge$\bigcirc$\hss}\hbox to 0pt{\hss $y$\hss}\hss}\\[2ex]
\hbox to 2em{\hss\hbox to 0pt{\hss\Huge$\bigcirc$\hss}\hbox to 0pt{\hss
{\it x}\hss}\hss}&  &\hbox to 2em{\hss\hbox to
0pt{\hss\Huge$\bigcirc$\hss}\hbox to 0pt{\hss {\it z}\hss}\hss}\\
r_1&  & r_1\\[-1.5ex]
& \nearrow \\
\hbox to 2em{\hss\hbox to 0pt{\hss\Huge$\square$\hss}\hbox to 0pt{\hss \raisebox{5pt}{\it z}\hss}\hss}
    & \mathop{\longrightarrow}
      &\hbox to 2em{\hss\hbox to 0pt{\hss\Huge$\square$\hss}\hbox to 0pt{\hss \raisebox{5pt}{\it z}\hss}\hss}\\
(k)&  & (k)\\[2ex]
&\mbox{\kern-20pt\rm \opcode{lw}~$r_1$~$k({\reg{sp}})$\kern-20pt}&\\
&\mbox{\kern-20pt\rm \opcode{get}~$r_1$~$k$\kern-20pt}&
\end{array}
&
\begin{array}[b]{ccc}
  {\reg{sp}}   & & {\reg{sp}}\\
 \hbox to 2em{\hss\hbox to 0pt{\hss\Huge$\bigcirc$\hss}\hbox to 0pt{\hss\raisebox{4pt}{$y\kern-1pt {+}\kern-1pt n$}\hss}\hss} & &\hbox to 2em{\hss\hbox to 0pt{\hss\Huge$\bigcirc$\hss}\hbox to 0pt{\hss\raisebox{4pt}{$y$}\hss}\hss}\\[-0.75ex]
  r_2& \kern-10pt{}^{\mbox{$n{\downarrow}$}}\kern-5pt\mathop{\nearrow}&r_2\\[-0.4ex]
\hbox to 2em{\hss\hbox to 0pt{\hss\Huge$\bigcirc$\hss}\hbox to 0pt{\hss\raisebox{4pt}{$y$}\hss}\hss}
  &\mathop{\longrightarrow}&
    \hbox to 2em{\hss\hbox to 0pt{\hss\Huge$\bigcirc$\hss}\hbox to 0pt{\hss\raisebox{4pt}{$y$}\hss}\hss} \\[2ex]
\hbox to 2em{\hss\hbox to 0pt{\hss\Huge$\square$\hss}\hss}&
\rdelim\}{4}{2.5em}[\begin{tabular}{l}$n/4$\\\hbox to
2.5em{finished\hss}\end{tabular}]&\\
 \kern-10pt (n{-}4)& &\\[0.25ex]
 \vbox to 2ex{\vss\vdots}& &\\
\hbox to 2em{\hss\hbox to
0pt{\hss\Huge$\square$\hss}\hss}
    &
       &\\
 (0) & &\\[2ex]
\multicolumn{3}{c}{\mbox{\rm \opcode{move}~{\reg{sp}}~$r_2$}}\\
\multicolumn{3}{c}{\mbox{\rm \opcode{rspf}~$r_2$}}
\end{array}
&
\begin{array}[b]{ccc}
r_3& & r_3\\
\hbox to 2em{\hss\hbox to 0pt{\hss\Huge$\bigcirc$\hss}\hbox to 0pt{\hss $z$\hss}\hss}& \mathop{\longrightarrow}&\hbox to 2em{\hss\hbox to 0pt{\hss\Huge$\bigcirc$\hss}\hbox to 0pt{\hss$z$\hss}\hss}\\
& \searrow^{\hbox to 0pt{$\type{c}^0$\hss}}\\[-2ex]
r_1& & r_1\\
\hbox to 2em{\hss\hbox to 0pt{\hss\Huge$\bigcirc$\hss}\hbox to 0pt{\hss
$x$\hss}\hss}& &\hbox to 2em{\hss\hbox to
0pt{\hss\Huge$\bigcirc$\hss}\hbox to
0pt{\hss\raisebox{2pt}{$y\overline{\mbox{\bf\&}}z$}\hss}\hss}\\[1ex]
& \nearrow_{\hbox to 0pt{\kern-2pt$\type{c}^0$\hss}}\\[-1ex]
\hbox to 2em{\hss\hbox to 0pt{\hss\Huge$\bigcirc$\hss}\hbox to 0pt{\hss $y$\hss}\hss}& \mathop{\longrightarrow}&\hbox to 2em{\hss\hbox to 0pt{\hss\Huge$\bigcirc$\hss}\hbox to 0pt{\hss\raisebox{2pt}{$y$}\hss}\hss}\\
r_2& & r_2\\[2ex]
\multicolumn{3}{c}{\mbox{\rm \opcode{nand}~$r_1$~$r_2$~$r_3$}}\\
\multicolumn{3}{c}{\mbox{\rm \opcode{nand}~$r_1$~$r_2$~$r_3$}}
\end{array}
\end{array}
\]
\caption{Dataflow semantics of four more machine code instructions.
}
\label{fig:5}
\end{figure}

Each trace that we consider ends with the \opcode{return} from a
subroutine call.  Only traces that have reached some register $r_2$ at that
moment are `properly terminated'. Any other trace (i.e., one
that has reached a stack cell) is not
considered further.  In the call protocol that we allow here,
the subroutine's local frame is
created at entry and destroyed at return and the data in it is not shared with
the caller:
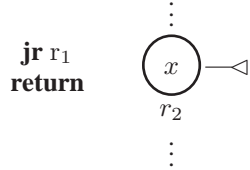
\[
\begin{array}{ccc}
\multicolumn{3}{c}{\rm \opcode{jr}~r_1}\\
\multicolumn{3}{c}{\rm \opcode{return}} \\[2ex]
\end{array}
\qquad
\begin{array}{ccc}
\vdots & &\\
  \hbox to 2em{\hss\hbox to 0pt{\hss\Huge$\bigcirc$\hss}\hbox to 0pt{\hss \raisebox{3pt}{$x$}\hss}\hss}
  &\raisebox{3.4pt}{\----}
  &\hskip-4pt\raisebox{3pt}{$\vartriangleleft$}
    \\
r_2 & & \\
\vdots & &
\end{array}
\]
We aim to constrain the possible sequences of events along traces.
The events are:
\begin{enumerate}
\item $!k$ for a write at stack offset $k$ with
$\opcode{put}~r~k$ (or \opcode{putx}, \opcode{swth} for strings, arrays);
\item $?k$ for a read at stack offset $k$ with
$\opcode{get}~r~k$ (or \opcode{getx}, \opcode{lwfh} for strings, arrays);
\item $\type{u}^n$ for the introduction of an array data address $a$ via
$\opcode{newh}~r~a~n$;
\item $\type{c}^n$ for the introduction of a `string' data address $a$ via
$\opcode{newx}~r~a~n$;
\item $\tau$ for the introduction of data of any kind $\tau$ `by hypothesis';
\item $\type{c}^0$ for the production of new data via
the \opcode{addaiu} or other arithmetic instruction;
\item $n{\uparrow}$ for the creation of a new stack frame of size
$n$ bytes via $\opcode{push}~n$;
\item $n{\downarrow}$ for restoring the previous stack frame,
terminating a frame of size $n$ bytes via $\opcode{rspf}~n$.
(or \opcode{stepx} when moving along a string);
\item nothing, for maintaining the data as-is or copying it. 
\end{enumerate}
An event does not always occur on the link one might expect: for
example, reading data to $r_1$ with \opcode{lw}~$r_1$~$4k(\reg{sp})$
evokes an event on a `\reg{sp} to \reg{sp}' link in Fig.~\ref{fig:5},
not on the `$(k)$ to $r_1$' (`stack slot $k$ to register $r_1$') link
that the data flows along.  We wish to enforce the following
restrictions. First, on the stack pointer:
\begin{figure}[tb]
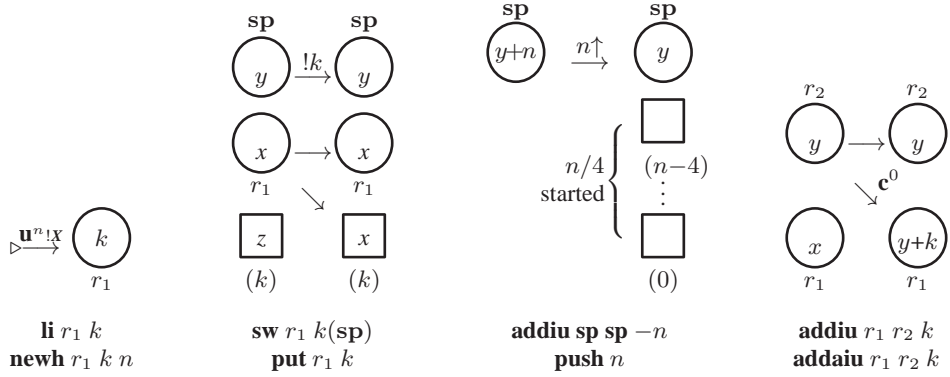

\[
\begin{array}{c@{\qquad\qquad}c@{\qquad\qquad}c@{\qquad\qquad}c}
\begin{array}[b]{ccc}
\triangleright\kern-4pt&
\mathop{\longrightarrow}\limits^{\mbox{\type{u}}^n{!}\typevar{X}} &\hbox to 2em{\hss\hbox to 0pt{\hss\Huge$\bigcirc$\hss}\hbox to 0pt{\hss \raisebox{3pt}{$k$}\hss}\hss}\\
& & r_1\\[2ex]
\multicolumn{3}{c}{\mbox{\rm \opcode{li}~$r_1$~$k$}}\\
\multicolumn{3}{c}{\mbox{\rm \opcode{newh}~$r_1$~$k$~$n$}}
\end{array}
&
\begin{array}[b]{ccc}
{\reg{sp}} &   & {\reg{sp}}\\[0.2ex]
\hbox to 2em{\hss\hbox to 0pt{\hss\Huge$\bigcirc$\hss}\hbox to 0pt{\hss $y$\hss}\hss}& \mathop{\longrightarrow}\limits^{\mbox{$!k$}} &\hbox to 2em{\hss\hbox to 0pt{\hss\Huge$\bigcirc$\hss}\hbox to 0pt{\hss $y$\hss}\hss}\\[2ex]
\hbox to 2em{\hss\hbox to 0pt{\hss\Huge$\bigcirc$\hss}\hbox to 0pt{\hss {\it x}\hss}\hss}& \mathop{\longrightarrow} &\hbox to 2em{\hss\hbox to 0pt{\hss\Huge$\bigcirc$\hss}\hbox to 0pt{\hss {\it x}\hss}\hss}\\
r_1&  & r_1\\[-1.5ex]
& \searrow \\
\hbox to 2em{\hss\hbox to 0pt{\hss\Huge$\square$\hss}\hbox to 0pt{\hss \raisebox{5pt}{\it z}\hss}\hss}
  &
    &\hbox to 2em{\hss\hbox to 0pt{\hss\Huge$\square$\hss}\hbox to 0pt{\hss \raisebox{5pt}{\it x}\hss}\hss}\\
(k)&  & (k)\\[2ex]
&\mbox{\kern-20pt\rm \opcode{sw}~$r_1$~$k({\reg{sp}})$\kern-20pt}&\\
&\mbox{\kern-20pt\rm \opcode{put}~$r_1$~$k$\kern-20pt}&
\end{array}
&
\begin{array}[b]{ccc}
{\reg{sp}}& & {\reg{sp}}\\
\hbox to 2em{\hss\hbox to 0pt{\hss\Huge$\bigcirc$\hss}\hbox to
0pt{\hss\raisebox{4pt}{$y{\kern-1pt+\kern-1pt}n$}\hss}\hss}&
\mathop{\longrightarrow}\limits^{\mbox{$n{\uparrow}$}} &\hbox to 2em{\hss\hbox to 0pt{\hss\Huge$\bigcirc$\hss}\hbox to 0pt{\hss\raisebox{4pt}{$y$}\hss}\hss}\\[2ex]
& \ldelim\{{4}{3.5em}[\begin{tabular}{@{}r@{}}$n/4$\\\hbox to 2.25em{\hss
started}\end{tabular}]&\hbox to 2em{\hss\hbox to 0pt{\hss\Huge$\square$\hss}\hss}\\
& & (n{-}4)\kern-10pt\\[0.25ex]
& & \vbox to 2ex{\vss\vdots}\\
& &\hbox to 2em{\hss\hbox to 0pt{\hss\Huge$\square$\hss}\hss}\\
& & (0)\\[2ex]
\multicolumn{3}{c}{\kern-20pt\mbox{\rm \opcode{addiu}~{\reg{sp}}~{\reg{sp}}~${-}n$}\kern-20pt}\\
\multicolumn{3}{c}{\kern-20pt\mbox{\rm \opcode{push}~$n$}\kern-20pt}
\end{array}
&
\begin{array}[b]{ccc}
r_2& & r_2\\
\hbox to 2em{\hss\hbox to 0pt{\hss\Huge$\bigcirc$\hss}\hbox to 0pt{\hss $y$\hss}\hss}& \mathop{\longrightarrow}&\hbox to 2em{\hss\hbox to 0pt{\hss\Huge$\bigcirc$\hss}\hbox to 0pt{\hss$y$\hss}\hss}\\[1ex]
& \searrow^{\hbox to 0pt{$\type{c}^0$\hss}}\\
\hbox to 2em{\hss\hbox to 0pt{\hss\Huge$\bigcirc$\hss}\hbox to 0pt{\hss
$x$\hss}\hss}& &\hbox to 2em{\hss\hbox to
0pt{\hss\Huge$\bigcirc$\hss}\hbox to 0pt{\hss\raisebox{2pt}{$y{\plus}k$}\hss}\hss}\\
r_1& & r_1\\[2ex]
\multicolumn{3}{c}{\mbox{\rm \opcode{addiu}~$r_1$~$r_2$~$k$}}\\
\multicolumn{3}{c}{\mbox{\rm \opcode{addaiu}~$r_1$~$r_2$~$k$}}\\
\end{array}
\end{array}
\]
\caption{Dataflow semantics of machine code/assembly language instructions.
}
\label{fig:4}
\end{figure}
%
%
\begin{enumerate}[(a)]
\item every $!k$ and $?k$ event is preceded by a last $n{\uparrow}$
event that has $n-w\ge k\ge 0$ (where $w$ is the number of bytes
written), so stack reads and writes do not step outside the local
frame of the subroutine;
\label{cons:1}
\item every $?k$ event is preceded by a $!k$ event that takes place
after the last preceding $n{\uparrow}$ event, so every read is of
something that has been written;
\item every $n{\downarrow}$ event is preceded by a last $m{\uparrow}$
event with $m=n$, and so on recursively 
so stack pushes and pops match up like parentheses;
\item no trace containing a \type{c} or \type{u} event other than an
originating $\type{c}^0$ may
eventually pass through the stack pointer register, so the only
operations allowed on the stack pointer are shifts up and down;
\item every $n{\uparrow}$ event is with $n> 0$.
\end{enumerate}
Secondly, on the traces through registers containing a string pointer:
\begin{enumerate}[(a)]
\item every $!k$ and $?k$ event is within the
bound $n$ established by the introduction $\type{c}^{\ddot{n}}$
on the trace, in that $n-w\ge k\ge 0$, where $w$ is the width of the
transferred data;
\item there is no (\ref{str:b}) constraint;
\label{str:b}
\item every $n{\downarrow}$ event is with $n$ equal
to the string increment established by the introduction
$\type{c}^{\ddot{n}}$ on the trace;
\item  no trace containing any other event than the
$\type{c}^{\ddot{n}}$ introduction and subsequent $n{\downarrow}$
shifts may later pass through the string pointer register, so
the only modifications allowed to the string pointer are shifts down;
\item there is no (\ref{str:e}) constraint.
\label{str:e}
\end{enumerate}
The constraints applied to traces through array pointers are stricter:
\begin{enumerate}[(a)]
\item every $!k$ and $?k$ event is within the
bound $n$ established by the preceding introduction $\type{u}^{n}$
on the trace, in that $n-w\ge k\ge 0$.
\item there is no (\ref{arr:b}) constraint;
\label{arr:b}
\item there are no  $n{\downarrow}$ or $n{\uparrow}$ events allowed;
\item no trace containing any other event than the
$\type{u}^{n}$ introduction may later pass through the array pointer
register, so no modifications to the array pointer are allowed;
\item there is no (\ref{arr:e}) constraint.
\label{arr:e}
\end{enumerate}
We express these constraints formally below. Starting with the
event that introduces an annotated type $\tau$ we accumulate a running
`total' annotated type along each trace.
The first two equations and their guards express the
constraints on an array pointer.  Shifts of the base address are not
allowed and reads and writes are restricted to the array bound:
\begin{align}
{\type{u}^n!\typevar{X}} \cdot {!}k      &= 
                 \type{u}^n!(\typevar{X}\cup\{k\})
                                           && n-w\ge k \ge 0
\label{eq:1}\\
{\type{u}^n!\typevar{X}} \cdot {?}k          &=
                  \type{u}^{n}!\typevar{X}
                                           &&  \typevar{X}\ni k\ge 0
\label{eq:2}
\intertext{The next three equations express the constraints
on a string pointer. Additionally, over the array pointer equations,
shifts-down on (increasing) the pointer are allowed:}
\type{c}^{\ddot{n}}!\typevar{X} \cdot n{\downarrow} &=
                 \type{c}^{\ddot{n}}   && n > 0
\label{eq:3}\\
{\type{c}^{\ddot{n}}!\typevar{X}} \cdot {!}k      &= 
                 \type{c}^{\ddot{n}}!(\typevar{X}\cup\{k\})
                                           && n-w\ge k \ge 0
\label{eq:4}\\
{\type{c}^{\ddot{n}}!\typevar{X}} \cdot {?}k          &=
                  \type{c}^{\ddot{n}}!\typevar{X}
                                           &&  \typevar{X}\ni k\ge 0
\label{eq:5}\\
\intertext{%
The next four equations 
express the constraints on the stack pointer.  Additionally, over the
string pointer equations,  shifts-up on (decreasing) the pointer are allowed.
The first two equations make shifts nest like parentheses:
}
\type{c}^f!\typevar{X} \cdot n{\uparrow}     &= \type{c}^{n^f}   && n > 0
\label{eq:6}\\
\type{c}^{n^{f}}!\typevar{X} \cdot n{\downarrow} &= \type{c}^f   && n > 0
\label{eq:7}\\
{\type{c}^{n^f}!\typevar{X}} \cdot {!}k      &= 
                 \type{c}^{n^f}!(\typevar{X}\cup\{k\})
                                           && n-w \ge k \ge 0
\label{eq:8}\\
{\type{c}^{n^f}!\typevar{X}} \cdot {?}k          &=
                 \type{c}^{n^f}!\typevar{X}
                                           &&  \typevar{X}\ni k\ge 0
\label{eq:9}
\end{align}
These calculations bind an annotated type to each register and stack
cell  at each point in the program.

Does the same register get the same type in every
trace calculation? Traces converge only after a \opcode{nand} (when
the type computed is $\type{c}^0$, so `yes it does' in this case) and
after a jump or branch. In these latter two cases we specify:
\begin{itemize}
\item[]%
The calculated type at the same registers or stack slots must be the same
across different traces starting from the same entry point for the programs
considered.\hfill(*)
\end{itemize}
The programs in which (*) is true are the only programs we consider.
They are programs that re-establish the same pattern of
annotated types at each point at every pass through a loop and no matter
which path through to a given point is taken. 

The annotated types that get bound to registers and stack slots are
the values in the states of an {\em abstract stack machine} whose
instruction semantics is described by Figs.~\ref{fig:4} and~\ref{fig:5}.
That may be shown to be an abstract
interpretation of the instruction trace semantics in a stack machine.
That in turn abstracts a machine code processor via disassembly.

Call an attempt in the stack machine to read or write beyond the current
local frame {\em out-of-bounds}.  That the abstract stack machine that
calculates with annotated types is an abstract interpretation of the
stack machine that calculates with integer words means that an
out-of-bounds access in the stack machine must evoke a $!k$ or $?k$
event on a trace through the abstract stack machine where $k$ is not
bounded by the size $n$ of the last $n{\uparrow}$ event on the trace.
But that is forbidden by (\ref{eq:1}-\ref{eq:9}) in the abstract stack
machine.  So if we can verify that (\ref{eq:1}-\ref{eq:9}) hold of a
program in the abstract stack machine, out-of-bounds accesses cannot
happen in the stack machine.

If out-of-bounds accesses in the stack machine cannot happen, then
we argue that aliasing cannot happen in the machine code processor.
The argument goes as follows:
the base address used for access via the RISC \opcode{lw} or
\opcode{sw} instructions must be either
\begin{enumerate}
\item the stack pointer (disassembly is to
\opcode{put}, \opcode{get}, \opcode{putb}, \opcode{getb} and the base
address register gets the annotated type $\type{c}^f!\typevar{X}$ for
some finite tower of frame sizes $f$);
\item the base address of a string, incremented several times by the
string increment (disassembly is to
\opcode{putx}, \opcode{getx}, \opcode{putbx}, \opcode{getbx} and the
base address register gets the annotated type
$\type{c}^{\ddot{n}}!\typevar{X}$ for some string step $n$);
\item the base address of an array (disassembly is to
\opcode{swth}, \opcode{sbth}, \opcode{lwfh}, \opcode{lbfh}
and the base address register gets the annotated type
$\type{u}^n!\typevar{X}$ for some array size $n$).
\end{enumerate}
Those are the only annotated types allowed by (\ref{eq:1}-\ref{eq:9})
on the abstract stack machine to be bound to the pointer's register at
the moment the event $!k$ or $?k$ happens.

In the first case, the offset in the accessing instruction is less than
the stack frame size, in the second case less than the string increment,
and in the third case less than the array size.  Those calculations are
the only ones that can be made for the address of the accessed element,
and they are each unique. For example, in case 1, the address used is
$s+k$, where $s$ is the stack pointer and $0\le k\le n-w$, where $n$ is
the local frame size and $w$ is the size of the data accessed. If two
such accesses from the same frame are at arithmetically equal address
aliases $s+k_1\equiv s+k_2$ but $s+k_1\ne s+k_2$ identically. So
$k_1\equiv k_2$ arithmetically but $k_1\ne k_2$ identically.  But $k_1$
and $k_2$ are small numbers in the range $0$ to $n$, where $n$ is the
frame size.  If they cannot be distinguished by the processor
arithmetic, then something is deeply wrong with the processor design.
Accessing an element of a parent frame with $s_1+k_1\equiv s_2+k_2$ where
$s_1=s_2-n$ is simply out of the question because $k_1$ is restricted to
the range $0$ to $n$.

We conclude that accessing different aliases of the same
address is impossible if the abstract interpretation of the program as
set out by Figs.~\ref{fig:4} and~\ref{fig:5} can be verified to satisfy
(\ref{eq:1}-\ref{eq:9}).

\end{document}